\documentclass[12pt]{article}
\usepackage[dvipdfmx]{graphicx}
\usepackage{amsfonts,amssymb,epsf}
\usepackage{amsmath,braket}
\usepackage{physics}
\usepackage{hyperref,color}
\usepackage{bigints}
\usepackage[toc,page]{appendix}
\usepackage{tikz}
\usepackage{ulem}
\usepackage{afterpage}
\usepackage{comment}
\hypersetup{
    unicode=false,          
    pdftoolbar=true,        
    pdfmenubar=true,        
    pdffitwindow=false,     
    pdfstartview={FitH},    
    pdfnewwindow=true,      
    colorlinks=true,       
    linkcolor=blue,          
    citecolor=blue,        
    filecolor=blue,      
    urlcolor=blue           
}
\usepackage[T1]{fontenc}
\usepackage[latin9]{inputenc}

\usepackage{braket}
\usepackage[hang,small,bf]{caption}
\usepackage[subrefformat=parens]{subcaption}
\usepackage{color}

\makeatletter
\numberwithin{figure}{section}
\numberwithin{equation}{section}
\numberwithin{table}{section}

\newcommand{\be}{\begin{equation}}
\newcommand{\ba}{\begin{eqnarray}}
\newcommand{\ea}{\end{eqnarray}}
\newcommand{\ee}{\end{equation}}

\newcommand{\bea}{\begin{eqnarray}}
\newcommand{\eea}{\end{eqnarray}}
\newcommand{\bes}{\begin{equation*}}
\newcommand{\beas}{\begin{eqnarray*}}
\newcommand{\eeas}{\end{eqnarray*}}
\newcommand{\bas}{\begin{array*}}
\newcommand{\eas}{\end{array*}}
\newcommand{\ees}{\end{equation*}}
\newcommand{\nn}{\nonumber}
\newcommand{\p}{\partial}

\makeatother
\allowdisplaybreaks
\begin{document}

\begin{titlepage}
\thispagestyle{empty}

\begin{flushright}
UT-Komaba/24-8, YITP-24-155
\\
\end{flushright}

\bigskip

\begin{center}
\noindent{\large \bf Holographic  Entanglement Entropy in the \! FLRW Universe}\\
\vspace{2cm}

Toshifumi Noumi$^{\dagger}$\footnote{
 tnoumi@g.ecc.u-tokyo.ac.jp}, Fumiya Sano$^{\flat,\natural}$\footnote{sanof.cosmo@gmail.com} and 
Yu-ki Suzuki$^\sharp$\footnote{yu-ki.suzuki@yukawa.kyoto-u.ac.jp}
\vspace{1cm}\\

{\it $^\dagger$ Graduate School of Arts and Sciences, The University of Tokyo, Tokyo 153-8902, Japan}\\

\vspace{0.1cm}
$^\flat$ \textit{Department of Physics, Institute of Science Tokyo, Tokyo 152-8551, Japan} \\
\vspace{0.1cm}
$^\natural$ \textit{Center for Theoretical Physics of the Universe, 
    Institute for Basic Science,\\Daejeon 34126, Korea}\\
\vspace{0.1cm}

{\it $^\sharp$Center for Gravitational Physics and Quantum Information,\\
Yukawa Institute for Theoretical Physics, Kyoto University, \\
Kitashirakawa Oiwakecho, Sakyo-ku, Kyoto 606-8502, Japan}

\setcounter{footnote}{0}
\vskip 2cm
\end{center}

\begin{center}\textbf{Abstract}\end{center}

We compute the holographic entanglement entropy via the Ryu--Takayanagi prescription in the three-dimensional Friedmann--Lema\^itre--Robertson--Walker universe. We consider two types of holographic scenarios analogous to the static patch holography and half de Sitter holography, in which the holographic boundary is timelike and placed in the bulk. We find in general that strong subadditivity can be satisfied only in the former scenario, and, in addition, the holographic boundary must lie within the apparent horizon. Also, for the universe filled with an ideal fluid with constant equation of state $w<-1$, the condition is sharpened as that the holographic boundary must lie within the event horizon instead. These constraints provide necessary conditions for the dual quantum field theory to be standard and compatible with strong subadditivity.

\end{titlepage}

\newpage

\tableofcontents
\newpage

\section{Introduction and summary}
After the discovery of the AdS/CFT (Anti-de Sitter/Conformal field theory) correspondence \cite{Maldacena:1997re}, theoretical physicists have been trying to understand the emergence of spacetime. One prominent way is to utilize quantum information theory in the AdS/CFT correspondence. The celebrated Ryu--Takayanagi  (RT) formula \cite{Ryu:2006bv,Ryu:2006ef} proposed a gravity dual of entanglement entropy that can be regarded as  a natural extension of Bekenstein--Hawking entropy \cite{Bekenstein:1973ur,Hawking:1975vcx}.  In the slogan of ``It from Qubit'', which states that the spacetime emerges through quantum entanglement, significant progress has been made in understanding quantum gravity in the AdS space.

\medskip
An extension of the holographic principle to the de Sitter (dS) spacetime, called the dS/CFT correspondence, was initiated by Strominger in Ref.~\cite{Strominger:2001pn}.  Although dS/CFT is not indicated from the string theory unlike AdS/CFT, various issues can be studied by assuming the would-be correspondence in a bottom-up spirit. For example, in the original dS/CFT correspondence \cite{Strominger:2001pn}, the dual CFT is on the future infinity of the de Sitter space and hence it lacks the time direction. The resulting CFT is expected to be non-unitary, which can also be expected from the analytic continuation of the radius from the AdS case \cite{Maldacena:2002vr}. 

\medskip
From the viewpoint of cosmology, the de Sitter space is directly related to our universe through cosmic inflation~\cite{Starobinsky:1980te,Sato:1980yn,Guth:1980zm,Linde:1981mu}, which is approximated by quasi-de Sitter space with broken time-translational symmetry due to the scalar field, inflaton, driving inflation. Correlation functions of the inflaton and graviton perturbations on the future infinity provide initial conditions for late-time anisotropy and inhomogeneity, which is observable through cosmic microwave background, large-scale structures, and so on (see~\cite{Planck:2018jri} for current constraints). Besides bulk analyses of inflationary perturbations, studies of inflationary correlators in the context of dS/CFT correspondence have also been conducted. Following pioneering works~\cite{Larsen:2002et,Maldacena:2002vr,vanderSchaar:2003sz,Seery:2006tq},
several phenomenological aspects have been explored such as operator product expansion for non-Gaussian inflationary correlators~\cite{Kehagias:2012pd}, correspondence of slow-roll inflation/deformed CFT~\cite{Bzowski:2012ih,Schalm:2012pi,Garriga:2013rpa}, initial vacuum states for wavefunction~\cite{Anninos:2014lwa,Shukla:2016bnu} etc., in addition to the so-called cosmological bootstrap~\cite{Arkani-Hamed:2018kmz,Baumann:2019oyu,Baumann:2020dch,Pajer:2020wnj}. However, due to unhealthy dual CFT, as well as the additional complexity from broken de Sitter isometry during inflation, the practical utility of dS/CFT correspondence is still limited to reproducing some simple results of quantum field theory (QFT) in (quasi-)dS space. 

\medskip
Instead, to realize a QFT dual with a time direction as in the AdS/CFT correspondence, we introduce a timelike holographic boundary (also referred to as a stretched horizon)  as proposed in Refs.~\cite{Susskind:2021omt,Kawamoto:2023nki} (see related works \cite{Anninos:2024wpy,Silverstein:2022dfj,Batra:2024kjl}). We impose the Dirichlet-like boundary conditions on the metric so that the dual QFT is decoupled from bulk gravitational modes.  In these cases, the bulk gravity is expected to be dual to a non-gravitating quantum field theory on the stretched horizon. The QFT may exhibit non-local behavior \cite{Kawamoto:2023nki} since it is a finite-cutoff holography as in the case of $T\bar{T}$-deformation \cite{McGough:2016lol,Guica:2019nzm} (also see \cite{Lewkowycz:2019xse} for a connection to the violation of the strong subadditivity of the holographic entanglement entropy \cite{Franken:2024wmh,Mori:2024gwe}), but it is believed to be unitary. In these cases, the holographic boundary is not located at the asymptotic boundary, and it is not clear where we should place the holographic boundary from first principles. 

\medskip
An extension of the holographic principle to other general (closed) Friedmann--Lema\^itre--Robertson--Walker (FLRW) spacetimes have been recently proposed in Refs.~\cite{Nomura:2016aww,Nomura:2016ikr,Franken:2023jas}, shedding light on the possibility that the entire history of our universe can be understood via holography and that the scientific reason why our universe is chosen from enormous possibilities can be analyzed.

\medskip
In this paper we address the following question: where should we place the holographic boundary and are there any holographic principles in the FLRW universes? We assume that the RT prescription is applicable in these spacetimes and explore constraints on holographic scenarios and background spacetimes by analyzing inequalities that the entanglement entropy should satisfy. More specifically,
we compute the holographic entanglement entropy via the RT formula in three-dimensional flat, closed, and open FLRW spacetimes and analyze whether the strong subadditivity (SSA) is satisfied. We also use the ideal fluid approximation for the matter and parametrize the FLRW spacetimes by $w$, which relates the energy density $\varepsilon$ and pressure $p$ as $p=w\varepsilon$. As a holographic scenario, we consider two types of location of the holographic boundary. One is referred to as half holography, corresponding a straight hypersurface\footnote{It is motivated by Ref.~\cite{Kawamoto:2023nki}, although the details are different.}, and the other is what we call a horizon-type holography, which is a circle-shaped hypersurface.  We define a subsystem on a constant time slice and test the SSA. We present explicit computations for the flat FLRW cases. For the half holography, the SSA is always violated. For the horizon-type holography,  the SSA is satisfied when the boundary is on or inside the apparent horizon and the event horizon for $w>-1$ and $w<-1$, respectively. In both choices of boundaries, the entanglement entropy exhibits a peculiar phase transition at the event horizon for the accelerating universe with $w<-1$.

\medskip
This paper is organized as follows. Sec.~\ref{pre} reviews holography for cosmologists and the flat FLRW universe for string theorists. We also summarize the relation between the SSA and the concavity of the entanglement entropy. Sec.~\ref{ds} presents the computation  in the flat de Sitter universe. In the de Sitter case, we can compute the geodesic distance by embedding into the Minkowski space and obtain the fully analytic results.
Sec.~\ref{flrwhol} extends the computation to the flat FLRW universe and obtain analytic results, where the integral of the geodesic can be evaluated analytically. 
Sec.~\ref{gee} comments on the matter contributions to the entropy. The paper concludes in Sec.~\ref{con}. 
For the closed and open universe, we review them in Appendix~\ref{app:open/closed} and conduct numerical computations of geodesic distances in the Appendix \ref{scale}. 

\section{Preliminaries} \label{pre}

In this section we provide a review of the background materials relevant to this paper.
Readers who are familiar with these topics may wish to skip this section.

\subsection{Review on holography}\label{hol}
The holographic principle was first proposed in Refs.~\cite{tHooft:1993dmi,Susskind:1994vu}.  A few years later, Maldacena proposed the AdS/CFT correspondence \cite{Maldacena:1997re} (see also \cite{Aharony:1999ti} for a comprehensive review) as an explicit realization. We begin by reviewing the AdS/CFT correspondence briefly and then introduce holography in de Sitter space. This section is intended for cosmologists who are new to the AdS/CFT correspondence. 

\subsubsection{AdS/CFT correspondence}
The AdS/CFT correspondence was originally proposed in the framework of string theory using D-branes. This type of proposal based on string theory are top-down approaches. On the other hand, the AdS/CFT correspondence is expected to be valid in a broader context and its general properties can be discussed without relying on concrete string theory realization, which is bottom-up approach. The statement of the AdS/CFT correspondence is that
\begin{align*}
    \text{\it Quantum gravity in $({d}{+}{1})$-dim AdS space is dual to $d$-dim QFT.}
\end{align*}
Rigorously, the QFT is CFT, which has a special conformal invariance in addition to a scale invariance, but the correspondence can be extended to the non-conformal cases by suitable deformations.
The AdS space has a timelike asymptotic boundary at spatial infinity, where the CFT lives. The metric for the global AdS$_{{d}{+}{1}}$ reads
\be
ds^2=-(1+r^2)dt^2+\frac{dr^2}{1+r^2}+r^2d\Omega_{d-1}^2\,,
\ee
where 
\be
0\leq r<\infty\,,\quad -\infty<t<\infty\,,
\ee
and $d\Omega_{d-1}^2$ is the line element on S$^{d-1}$. In this case the topology of boundary is $\mathbf{R}\times S^{d-1}$, but we can realize the correspondence in other topologies by taking other coordinates.  Note that the CFT is on $r=\infty$. 
The consistency of the AdS/CFT correspondence can be supported by the symmetry. The $({d}{+}{1})$-dim AdS space can be embedded into the flat spacetime, which admits two time directions, and respects a conformal symmetry in $d$ dimensions. Both sides of the duality share the symmetry group $SO(2,d)$.

\medskip
In the subsequent papers \cite{Gubser:1998bc,Witten:1998qj} (see also \cite{Banks:1998dd} for a relevant work), the authors formulated the way to relate partition functions and correlation functions between gravity and QFT sides. The duality in the AdS/CFT correspondence states the isomorphism between the Hilbert spaces of both sides. In other words, there exists a one-to-one correspondence between the bulk (gravity side) and the boundary (QFT side) operators. For example, the gauge current and the stress-energy tensor in the boundary CFT correspond to the gauge field and metric in the bulk gravity. The emergent direction, present in the gravity side but absent in the QFT side, can be regarded as an energy scale through holographic renormalization \cite{Skenderis:2002wp,Bianchi:2001kw}. It is remarkable that the theories in different dimensions match, but roughly speaking, the interactions in the CFT side encoding the information of the emergent (radial) direction is complicated. 

\medskip
As a final remark, the quantum field theory discussed above was originally non-gravitating. However, there have been numerous attempts to realize a finite cutoff boundary, i.e., one not located at the asymptotic boundary, such as $T\bar{T}$-deformation \cite{Zamolodchikov:2004ce,McGough:2016lol}. Indeed, in this paper we deal with the cases of finite cutoff as boundary duals.

\subsubsection{Holography in the de Sitter space}

\subsubsection*{dS/CFT correpondence}

The first attempt to extend the AdS/CFT correspondence to the de Sitter case was initiated in Ref.~\cite{Strominger:2001pn} by Strominger. The dS space has a spacelike asymptotic boundary at the future infinity unlike the AdS case. The metric for the flat slicing of dS$_{{d}{+}{1}}$ reads
\be
ds^2=\frac{-d\eta^2+d\Vec{x}^2}{\eta^2}\,,
\ee
where $-\infty<\eta<0$. The CFT is realized on $\eta=0$ time slice.
The dual QFT (CFT) is expected to be non-unitary since a time direction is absent in the theory as can be inferred from the metric. This can also be understood by analytically continuing the radius of curvature from the AdS to dS first observed in \cite{Maldacena:2002vr}. In particular, the central charge of CFT$_2$ in the dS$_3$/CFT$_2$ is imaginary, i.e., $c=\frac{3iL}{2G}$ where $L$ is the curvature of radius and $G$ is the Newton constant, whereas the AdS$_3$/CFT$_2$ involves a real central charge. 
Furthermore, the emergent direction in the dS/CFT is the time direction. The dS/CFT requires gravity to play the role of time. 
In the paper \cite{Maldacena:2002vr}, a relation between the CFT partition function and the wavefunction of the universe is proposed as in the AdS/CFT case. Through this relation, we may compute $n$-point correlation functions by taking functional derivatives of the generating function. 
Note that unlike the AdS/CFT, it is not easy to realize a stable de Sitter vacua in string theory in a well-controlled manner (see \cite{Kachru:2003aw} for a candidate). 

\medskip
As a phenomenologically important example, the cosmic inflation is described by the quasi-de Sitter spacetime. Furthermore, the asymptotic boundary at future infinity effectively characterizes the end of inflation, even though the inflation terminates at a finite time in reality, giving rise to Big Bang universe. Thus, de Sitter holography would be helpful to understand the dawn of our universe through quantum gravity. 

\subsubsection*{Static patch holography}
The above dS/CFT proposal faces a fundamental issue since the boundary CFT does not have a time direction, leading to its non-unitarity. To resolve this problem, in Ref.~\cite{Susskind:2021omt}, Susskind proposed the static patch holography, which states that gravity in the static patch is dual to a QFT on the stretched horizon defined by some hypersurface. To describe that, let us introduce coordinates. The metric for the $({d}{+}{1})$-dim static patch is given by
\be
ds^2=-(1-r^2)dt^2+\frac{dr^2}{1-r^2}+r^2d\Omega_{d-1}^2\,,
\ee
where 
\be
0\leq r<1,\quad -\infty<t<\infty\,.
\ee
Since the static patch is a compact space and there is no timelike boundary originally. A possible candidate would be a timelike boundary defined by $r=r_*$. In particular, Ref.~\cite{Susskind:2021omt} considers a cosmological horizon $r=1$ as the boundary. However, it remains unsolved neither a prescription for deriving the correlation functions on both sides nor what kind of theories there are. One of the motivations of this paper is to provide a criteria for putting boundary in the static patch. More generally, we also extend this setup to the FLRW cases. It should be noted that, to obtain a non-gravitating QFT dual on the boundary, we should impose a Dirichlet-like boundary condition on the metric fluctuation.

\subsubsection*{A half de Sitter holography}
As another approach to resolve the non-unitarity of CFTs dual to dS gravity, Ref.~\cite{Kawamoto:2023nki} introduced a timelike boundary at a finite radial cutoff. They consider the global patch of the de Sitter space
\be
ds^2=-dt^2+\cosh^2 t (d\theta^2+\sin^2\theta d\phi^2)\,,
\ee
and introduce a timelike boundary at $\theta=\theta_*$, where the Dirichlet boundary condition is imposed on the metric.  Although the precise nature of the dual theory remains unclear, they expect it to be a non-local theory that possibly involves  infinitely higher derivative kinetic terms. This expectation is supported by the super-volume law of the holographic entanglement entropy. Our setup is a generalization of this setup, although they analyze the problem in the global  dS.

\subsection{Ryu--Takayanagi formula and strong subadditivity}
\label{subsec:RT}
In this section, we review the RT formula for the holographic entanglement entropy in the AdS gravity, which is dual to the entanglement entropy in the boundary CFT. We also explain SSA, which the entanglement entropy is expected to satisfy in the standard QFT.

\subsubsection{Ryu--Takayanagi formula}
In this paper, we utilize the holographic entanglement entropy via RT prescription to probe spacetimes. The computation of entanglement entropy in the CFT side was established in Ref.~\cite{Calabrese:2004eu} (see \cite{Holzhey:1994we} for an earlier related work). 
Then, Ryu and Takayanagi found in Ref.~\cite{Ryu:2006bv,Ryu:2006ef} that a gravity dual of the entanglement entropy, called holographic entanglement entropy, is identified with the area of the minimal surface connecting the boundaries of the subsystem, divided by $4G$. Note that we choose the minimal surface homologous to the subsystem, which plays a role in the non-trivial geometry like a thermal geometry. This can be regarded as a natural extension of the Bekenstein--Hawking entropy \cite{Bekenstein:1973ur,Hawking:1975vcx} since the entanglement entropy quantifies the information loss after tracing out a subsystem. Later, this holographic entanglement entropy was derived from gravitational path integral by Lewkowycz and Maldacena in \cite{Lewkowycz:2013nqa}.

\medskip
In this paper, we compute the holographic entanglement entropy using RT formula. Although it is not yet fully understood whether the RT formula can be applied to other spacetimes than the AdS space, we discuss what types of holographic scenarios can admit a standard QFT dual by assuming that the RT formula is applicable to FLRW spacetimes and requiring the SSA as a consistency condition. Note that, for simplicity of the calculation, we will focus on three dimensions in which the minimal surface reduces to the geodesic connecting the endpoints of the subsystem.

\subsubsection{Strong subadditivity}

It is natural to expect that the holographic entanglement entropy satisfies various inequalities for quantum entropy (see~\cite{Rangamani:2016dms} for a comprehensive review). In particular, we focus on the strong subadditivity~\cite{Lieb:1973zz,Lieb:1973cp}, abbreviated as SSA, which is related to the concavity of the entropy. The definition of the SSA is as follows. Let $A$, $B$, and $C$ be subsystems. Then, the SSA states that
\be
    S_{AB}+S_{BC}-S_{ABC}-S_B\geq0\,,\label{SSA}
\ee
where $S_A$ is the entanglement entropy of the subsystem $A$, and $AB$ is defined by $A\cup B$. 
For convenience, we consider subsystems on a constant time slice, which is often called the static SSA in the context of the AdS/CFT correspondence\footnote{It would also be interesting to analyze the boosted SSA (see \cite{Casini:2022rlv} for a review), where subsystems are Lorentz boosted and thus time dependent. It may provide stronger constraints than the static SSA.}. Since we suppose that the total spacetime dimension is three, the subsystem is one-dimensional. While it is intriguing to verify this inequality directly, we instead compute the second derivative with respect to the system size. Let the length of $ABC$ and $B$ be $x_1$ and $x_2$, respectively, and assume the symmetric configuration $AB=BC$. Then, we simplify (\ref{SSA}) as
\be
    2S\left(\frac{x_1+x_2}{2}\right)-S(x_1)-S(x_2)\geq0\,,
\ee
where we assumed that the system is homogeneous.
Finally, by setting $x_1=x+\Delta x$ and $x_2=x-\Delta x$ and taking $\Delta x\rightarrow0$, we obtain 
\be
    \frac{\p^2 S}{\p x^2}\leq 0\,.
\ee
For holographic entanglement entropy, a beautiful geometric proof of the SSA is provided in Ref.~\cite{Headrick:2007km} (see also \cite{Hirata:2006jx} for an earlier related study).

\subsection{Basics of the flat FLRW universe}\label{FLRW}

In this subsection, we review basics of the flat FLRW universe. The main text focuses on the flat case, while the closed and open cases are discussed in Appendix~\ref{scale}.

\paragraph{Geometry.}

Consider a flat FLRW universe in $({d}{+}{1})$-dimensions:
\begin{align}
ds^2=a^2(\eta)
\left[-d\eta^2+dr^2+r^2d\Omega_{d-1}^2\right]\,,
\end{align}
where $\eta$ is the conformal time, $a(\eta)$ is the scale factor of the universe, and $d\Omega_{d-1}^2$ is the line element of the $(d-1)$-dimensional unit sphere. While our analysis mainly uses the conformal time $\eta$, physical properties of the universe are often captured by the physical time $t$ defined as $dt=a(\eta)d\eta$, in terms of which the metric reads
\begin{align}
ds^2=-dt^2+a^2(\eta(t))
\left[dr^2+r^2d\Omega_{d-1}^2\right]\,.
\end{align}
For instance, the Hubble parameter $H=\dot{a}/a$, where a dot denotes differentiation with respect to physical time $t$, describes the expansion rate of the universe. Also, the acceleration of the universe is given by $\ddot{a}$.

\paragraph{Apparent horizon.}

In our holographic analysis, the cosmological horizon plays a crucial role. In the flat FLRW universe, the apparent horizon corresponds to a sphere expanding at the speed of light. See Appendix~\ref{scale} for the precise definition. Then, according to the Hubble law, the physical radius $R_{\rm H}$ of the apparent horizon is
\begin{align}
R_\mathrm{H}=H^{-1}\,.
\end{align}
Correspondingly, the location of the apparent horizon in terms of the comoving radius $r$ is specified as
\begin{align}
r=r_{\rm H}
\quad{\rm with}\quad
r_{\rm H}=\frac{1}{aH}=\frac{a}{a'}\,,
\end{align}
where a prime denotes a derivative in conformal time $\eta$.

\paragraph{Benchmark examples.}

For illustration, it is often useful to consider the universe filled with an ideal fluid of the constant equation of state $w=p/\varepsilon$, where $\varepsilon>0$ and $p$ are the energy density and pressure of the fluid, respectively. From the Einstein equations, time evolution of the scale factor follows as
\begin{align}
\label{power-law_genreal}
a(\eta)=R\left(\frac{\eta}{\gamma}\right)^\gamma
\quad
{\rm with}
\quad
\gamma=\frac{2}{d(1+w)-2}\,,
\end{align}
where $R$ is a constant that characterizes the physical size of the universe. For simplicity, we employ a normalization $R=1$ in the following. Also, we choose the expansion phase of the universe without loss of generality, so that $0<\eta<\infty$ for $\gamma>0$ and $-\infty<\eta<0$ for $\gamma<0$. Note that $\gamma>0$ and $\gamma<0$ correspond to the decelerating universe $\ddot{a}<0$ and the accelerating universe $\ddot{a}>0$, respectively. Particularly, the flat chart of de Sitter spacetime is reproduced by $w=-1$, or equivalently $\gamma=-1$, which saturates the null energy condition $w\geq-1$.\footnote{
Note that the null energy condition and the weak energy condition are equivalent in our setup because we assume the positive energy density $\varepsilon>0$.}
For the power-law scale factor~\eqref{power-law_genreal}, the comoving radius of the apparent horizon reads
\begin{align}
r_{\rm H}=\frac{\eta}{\gamma}\,.
\end{align}

\paragraph{Summary for three-dimensional universe.}

Let us rewrite the above equations specifically for the three-dimensional universe $d=2$, which is the setting of our holographic analysis. The metric in the conformal time is
\begin{align}
ds^2=a^2(\eta)
\left[
-d\eta^2+dr^2+r^2d\phi^2
\right]\,,
\end{align}
where $\phi$ is the $S^1$ angular coordinate with domain $0\leq\phi<2\pi$. For the flat FLRW universe with a constant equation of state parameter $w$, the parameter $\gamma$ is simply $\gamma=w^{-1}$, thus the scale factor and the comoving radius of the apparent horizon are
\begin{align}
a(\eta)=(w\eta)^{\frac{1}{w}}
\,,
\quad
r_{\rm H}=\frac{a}{a'}=w\eta\,.
\end{align}

\section{de Sitter holography}
\label{ds}

As the simplest example of the FLRW universe, we first analyze the de Sitter spacetime. The advantage of this case is that the length of the geodesic curve is derived analytically by embedding the spacetime into the Minkowski space.

\medskip
Three-dimensional de Sitter spacetime with the unit radius is defined by the embedding,
\be
-(X^0)^2+(X^1)^2+(X^2)^2+(X^3)^2=1\,,
\ee
into four-dimensional flat spacetime,
\be
ds^2=-(dX^0)^2+(dX^1)^2+(dX^2)^2+(dX^3)^2\,.
\ee
Under the embedding, the geodesic distance $D>0$ between two spatially separated points $X_A$ and $X_B$ is given by
  \be
  \label{geodesic_length_dS_emb}
  \cos D=-X_A^0 X_B^0+X_A^1 X_B^1+X_A^2 X_B^2+X_A^3 X_B^3\,,
  \ee
which provides a simple formula for the holographic entanglement entropy. In this section we consider the flat chart of de Sitter spacetime and examine the SSA in two types of holographic scenarios analogous to the static patch holography~\cite{Susskind:2021omt} and the half de Sitter holography~\cite{Kawamoto:2023nki}. For extension to the closed and open universes, see Appendix~\ref{app:open/closed}.

\subsection{Holographic scenarios}
\label{subsec:setup}

\paragraph{Flat universe.}

The flat chart of de Sitter spacetime is defined by
\begin{align}
    X^0=\frac{\eta^2-1-r^2}{2\eta},\quad X^1=\frac{\eta^2+1-r^2}{2\eta},\quad X^2=\frac{r\cos\phi}{\eta},\quad     X^3=\frac{r\sin\phi}{\eta}
\end{align}
with the coordinate domain
\be
-\infty<\eta<0\,,
\quad
0\leq r<\infty\,,
\quad 0\leq\phi<2\pi\,.
 \ee
The corresponding metric is given by
\be
ds^2=\frac{-d\eta^2+dr^2+r^2 d\phi^2}{\eta^2}\,.
\ee
The cosmological event horizon for the observer at $r=0$ is located at $r=|\eta|$. Furthermore, the geodesic distance $D$ of two points $(\eta_A,r_A,\phi_A)$ and $(\eta_B,r_B,\phi_B)$ reads
\be
\label{geodesic_dS_flat}
\cos D=
1-\frac{-(\eta_A-\eta_B)^2+r_A^2+r_B^2-2r_Ar_B\cos(\phi_A-\phi_B)}{2\eta_A\eta_B}
\,.
\ee

\paragraph{Holographic setups.}

As reviewed in Sec.~\ref{pre}, holographic scenarios for expanding universes can be classified into two types. In holographic scenarios with a spacelike boundary such as the dS/CFT correspondence~\cite{Strominger:2001pn}, the bulk and boundary do not share the notion of time and thus the corresponding holographic dual would be non-unitary. On the other hand, in scenarios with a timelike boundary such as the static patch holography~\cite{Susskind:2021omt} and the half de Sitter holography~\cite{Kawamoto:2023nki}, the bulk and boundary share the time and so the would-be holographic dual is expected to be unitary. Since the purpose of the current paper is to identify holographic scenarios with the standard holographic dual, we primarily focus on the latter class of setups, even though it is applicable to the former too. See also the last paragraph of the subsection.

\begin{figure}[t]
  \centering
    \includegraphics[keepaspectratio, width=0.65\linewidth]{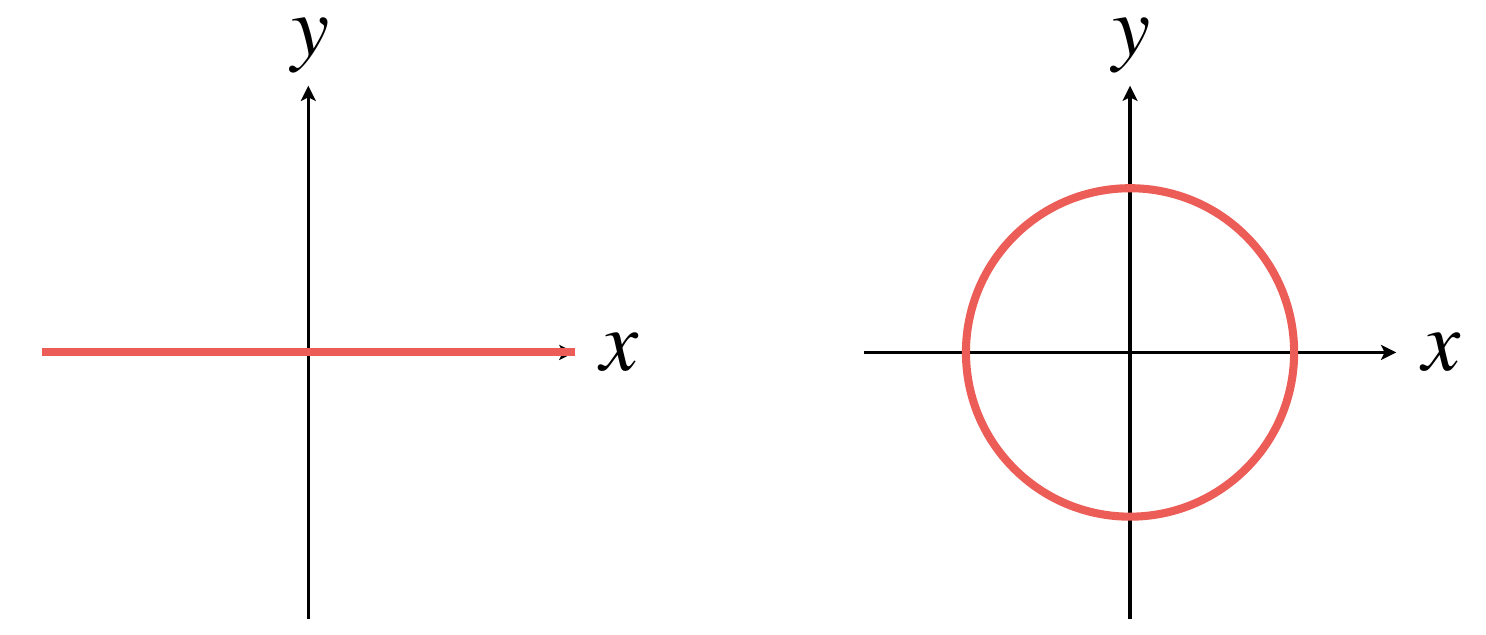}
\caption{Location of the boundary on the constant time slice is depicted by the red line for the half holography (left) and the horizon-type holography (right). }
\label{fig:screen}
\end{figure}

\medskip
With this motivation, we consider holographic scenarios with a $({1}{+}{1})$-dimensional timelike boundary, which extends along the conformal time ($\eta$) direction and a spatial direction. At a constant time $\eta=\eta_*$, the boundary is specified by embedding a one-dimensional curve into the $(r,\phi)$ plane. In this paper, we study the following two types of symmetric embedding (see also Fig.~\ref{fig:screen}):
\begin{enumerate}
\item Half holography.

The first type is motivated by the half de Sitter holography~\cite{Kawamoto:2023nki} where the boundary is placed at a time $\eta=\eta_*$ along a straight line defined by
\begin{align}
y=0
\quad
{\rm with}
\quad
(x,y):=(r\cos\phi,r\sin\phi)\,,
\end{align}
which cuts a constant-$\eta$ slice in half. We refer to this type as the half holography in the following. Thanks to the translational invariance of the flat universe, the boundary enjoys a shift symmetry regarding $x$ and therefore the second derivative of the entanglement entropy can be used to test the (strong) subadditivity.

\item Horizon-type holography.

The second type is motivated by the static patch holography~\cite{Susskind:2021omt} where the boundary is placed at a time $\eta=\eta_*$ along a circle defined by
\begin{align}
r=\lambda|\eta_*|\,,
\quad
0\leq\phi<2\pi\,.
\end{align}
Here, a parameter $\lambda$ is introduced for later convenience. The boundary coincides with the cosmological horizon when $\lambda=1$. On the other hand, for $0< \lambda<1$ and $\lambda>1$, the boundary is located inside and outside of the horizon, respectively. We refer to this type as the horizon-type holography in the following. Thanks to the rotational invariance of the flat universe, the boundary enjoys a shift symmetry regarding $\phi$ and therefore the second derivative of the entanglement entropy can be used to test the (strong) subadditivity.

\end{enumerate}

\paragraph{Remarks on holography with a spacelike boundary.}

As mentioned above, our analysis is also applicable to holographic scenarios with a spacelike boundary. For instance, suppose that a boundary is located on a spacelike surface defined by $\eta=\eta_*$, similarly to the dS/CFT correspondence~\cite{Strominger:2001pn}. If we identify the $y$ coordinate with a Euclidean time coordinate, the constant (Euclidean) time slice corresponds to that of the half holography. On the other hand, if we identify the radial coordinate $r$ with a Euclidean time, the constant time slice coincides with that of the horizon-type holography. Since the holographic entanglement entropy depends only on the boundary points of the subsystem at a given time, the following analysis can be applied to these scenarios without any modification.\footnote{Although the RT-prescription seems to yields the same answer as the timelike boundary case, the notion of the state would be different since the direction of the path integral is different. It is interesting to investigate this distinction in more detail.} This observation is also applicable to our analysis in open and closed universes of dS holography and the FLRW holography.

\subsection{Half holography}
\label{subsec:half_dS}

\begin{figure}[t]
  \centering
    \includegraphics[keepaspectratio, width=0.9\linewidth]{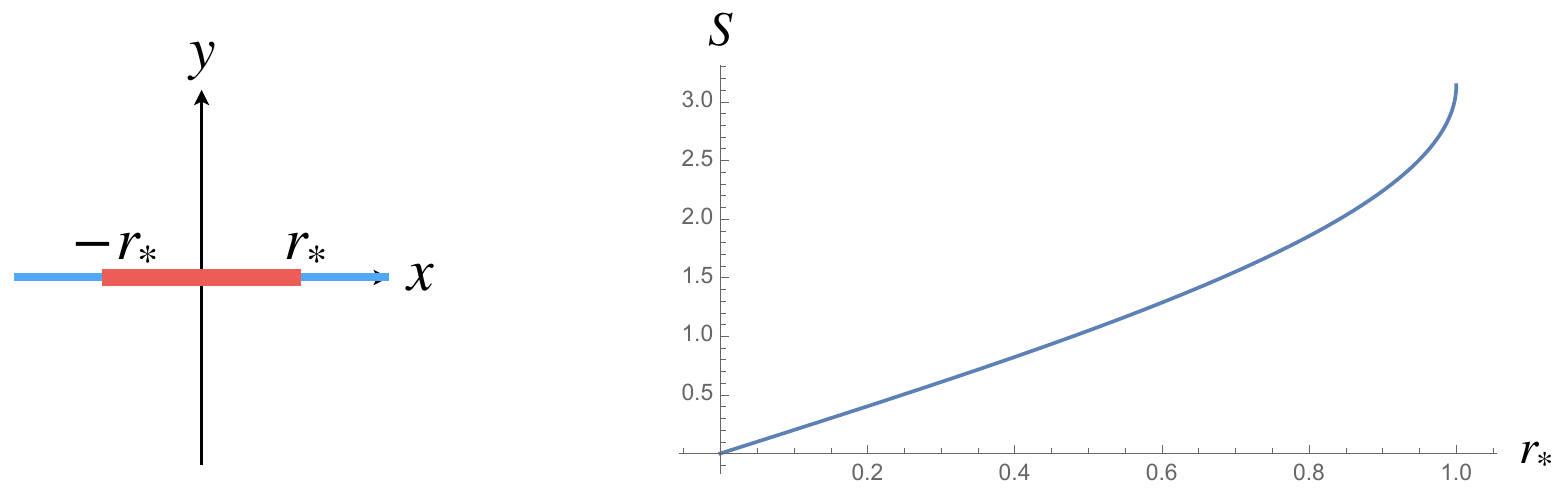}
\caption{Left: Subsystem (red) and its complement (blue) in the half holography.
\\
Right: Holographic entanglement entropy in the unit $4G=1$.}
\label{fig:half_dS}
\end{figure}

We now compute the holographic entanglement entropy for the half holography. We consider a subsystem on the interval defined by
\begin{align}
x\in (-r_*,r_*)\,,
\quad
y=0\,.
\end{align}
See also Fig.~\ref{fig:half_dS}. Without loss of generality, we choose the conformal time as $\eta_*=-1$ for visual clarity, using the dilatation symmetry of de Sitter space. The holographic entanglement entropy then follows from Eq.~\eqref{geodesic_dS_flat} and the RT formula as
\begin{align}
\label{half-flat-dS}
S=\frac{1}{2G}\arcsin r_*\,,
\end{align}
which is real as long as the subsystem interval lies inside the cosmological horizon, i.e., $r_*\leq1$. However, it becomes complex once the interval extends beyond the event horizon.\footnote{In the de Sitter space, the event horizon coincides with the apparent horizon.} $r_*>1$~\cite{Kawamoto:2023nki} The second derivative with respect to (half of) the subsystem size $r_*$ reads
\begin{align}
\frac{\partial^2S}{\partial r_*^2}=
\frac{1}{2G}\frac{r_*}{(1-r_*^2)^{\frac{3}{2}}}\,,
\end{align}
which is always positive as long as the subsystem interval is inside the event horizon. Hence, the half holography scenario violates the SSA. See also Fig.~\ref{fig:half_dS} for plots of the holographic entanglement entropy and its second derivative.

\subsection{Horizon-type holography}
\label{subsec:horizon-like_dS}

\begin{figure}[t]
  \centering
    \includegraphics[keepaspectratio, width=0.9\linewidth]{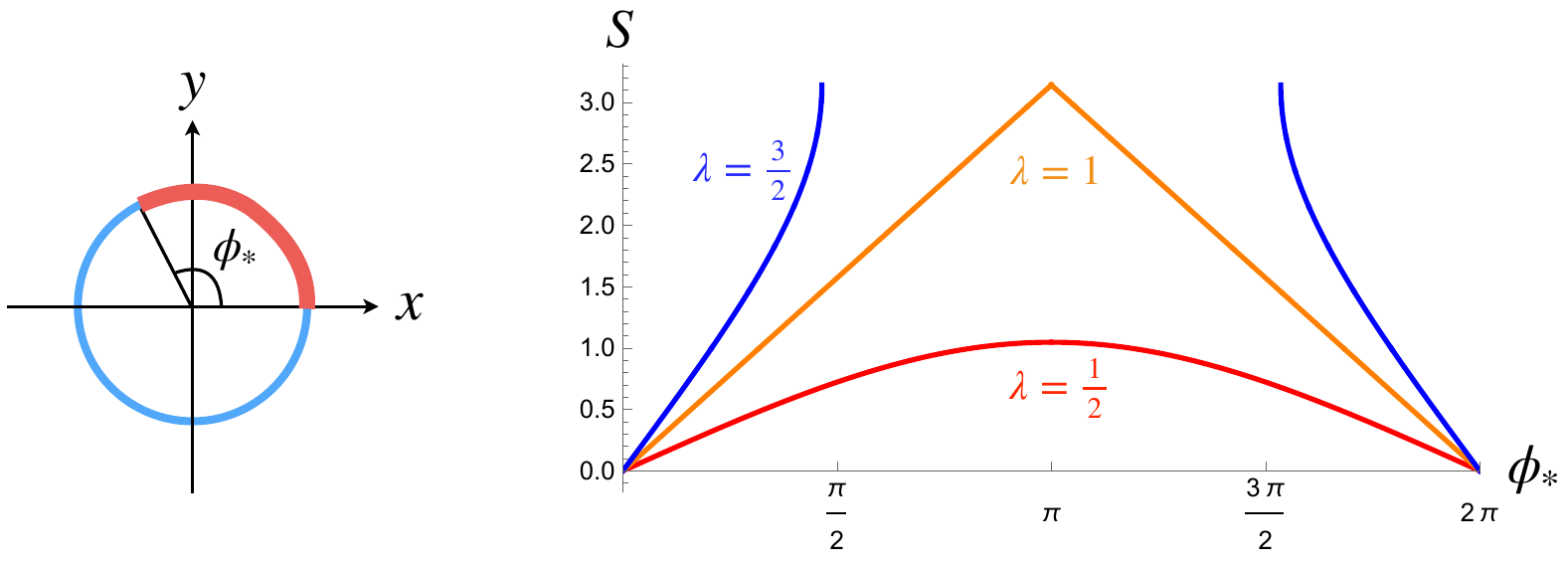}
\caption{Left: Subsystem (red) and its complement (blue) in the horizon-type holography. Right: Holographic entanglement entropy for $\lambda=\frac{1}{2},1,\frac{3}{2}$ in the unit $4G=1$.}
\label{fig:horizon_dS}
\end{figure}

Next we study the horizon-type holography. We consider a subsystem on an arc defined by
\begin{align}
r=\lambda|\eta_*|\,,
\quad
\phi\in(0,\phi_*)
\,.
\end{align}
See also Fig.~\ref{fig:horizon_dS}. As before, we choose $\eta_*=-1$ without loss of generality. Also, recall that $\lambda$ quantifies the boundary size relative to the horizon size. Then, the holographic entanglement entropy follows from Eq.~\eqref{geodesic_dS_flat}  as
\begin{align}
S=\frac{1}{4G}\arccos\left[
1-2\lambda^2\sin^2\frac{\phi_*}{2}
\right]\,,
\end{align}
which is real for all $\phi_*$ as long as the boundary is on or inside the horizon, i.e., $\lambda\leq1$. In contrast, if the boundary is outside of the horizon $\lambda>1$, the entropy becomes complex for
$\sin\frac{\phi_*}{2}>\frac{1}{\lambda}$.
On the other hand, the second derivative of the holographic entanglement entropy with respect to the subsystem size $\phi_*$ reads
\begin{align}
\frac{\partial^2 S}{\partial\phi_*^2}
=-\frac{(1-\lambda^2)}{8G}\frac{ \lambda\sin\frac{\phi_*}{2}}{(1-\lambda^2\sin^2\frac{\phi_*}{2})^{\frac{3}{2}}}\,.
\end{align}
Interestingly, it is negative for all $\phi_*$ when the boundary is inside the cosmological horizon, whereas it is imaginary when the boundary is outside the horizon, originally observed in \cite{Kawamoto:2023nki}. We thus conclude that the horizon-type holography  is compatible with the SSA as long as the boundary is on or inside the horizon, at least within the scope of our present analysis. On the other hand, placing the boundary outside of the horizon leads to a violation of SSA. See also Fig.~\ref{fig:horizon_dS}.

\bigskip
We comment on the case of $\lambda=1$, where the boundary lies on the cosmological horizon. For this, the holographic entanglement entropy is simply proportional to the size of the subsystem or its complement:
\begin{align}
S=\left\{\begin{array}{cl}
\displaystyle
\frac{\phi_*}{4G}
& (0<\phi_*<\pi)\,,
\\[3mm]
\displaystyle
\frac{2\pi-\phi_*}{4G}
& (\pi<\phi_*<2\pi)\,.\end{array}\right.
\end{align}
Hence, its second derivative vanishes except for $\phi_*=\pi$. It is suggestive that when the cosmological horizon is identified with the boundary, the negativity bound on the second derivative is saturated by the de Sitter spacetime, which also saturates the null energy condition. This motivates us to extend our analysis to more general FLRW universes and explore the role of the null energy condition.

\section{FLRW holography}\label{flrwhol}

In this section we extend the previous analysis in the de Sitter space to general FLRW universes. In Secs.~\ref{subsec:cosmo_setup}--\ref{FLRW:horizon}, we perform an analytic study of the holographic entanglement entropy in the flat universe filled with an ideal fluid satisfying a constant equation of state. In Sec.~\ref{subsec:where}, we conclude the analysis by discussing the possible location of the holographic boundary compatible with the SSA. See also Appendix~\ref{app:open/closed} for the corresponding analysis in closed and open universes.

\subsection{Cosmological setup}
\label{subsec:cosmo_setup}

\paragraph{Generality.}

Consider a flat FLRW universe in three dimensions:
\begin{align}
ds^2=a(\eta)^2\left[-d\eta^2+dr^2+r^2d\phi^2\right]
\qquad
(0\leq r<\infty\,,\,\, 0\leq\phi< 2\pi)
\,,
\end{align}
where the scale factor $a(\eta)$ is left arbitrary for this moment. In the following analysis of the holographic entanglement entropy, we need to evaluate the geodesic distance between two points $A,B$ on the same constant-$\eta$ slice $\eta=\eta_*$:
\begin{align}
(\eta,r,\phi)=(\eta_*,r_A,\phi_A),\,\,(\eta_*,r_B,\phi_B)\,.
\end{align}
For this purpose, it is convenient to use the translation and rotation symmetries of the flat slice $\eta=\eta_*$ to map the two points to
\begin{align}
\label{after_rotation}
(\eta,r,\phi)=(\eta_*,\tilde{r},0)\,,\,\,(\eta_*,\tilde{r},\pi)
\quad
{\rm with}
\quad
\tilde{r}=\frac{1}{2}\sqrt{r_A^2+r_B^2-2r_Ar_B\cos(\phi_A-\phi_B)}\,,
\end{align}
or equivalently, in the Euclidean coordinates $(x,y)=(r\cos\theta,r\sin\theta)$,
\begin{align}
\label{after_rotation2}
(\eta,x,y)=(\eta_*,\tilde{r},0)\,,\,\,(\eta_*,-\tilde{r},0)\,.
\end{align}
The symmetry of the problem implies that $y=0$ along the geodesic, so the geodesic is characterized by the conformal time $\eta(x)$ as a function of $x$. Then, the geodesic connecting the two points~\eqref{after_rotation2} is determined by minimizing the length,
\be
D(\eta_*,\tilde{r})=\int_{-\tilde{r}}^{\tilde{r}}dx\, a\big(\eta(x)\big) \sqrt{1-\eta'(x)^2}=2\int_{0}^{\tilde{r}}dx\, a\big(\eta(x)\big) \sqrt{1-\eta'(x)^2}\,,
\label{eq:geodesic}
\ee
under the boundary conditions $\eta'(0)=0$ and $\eta(\tilde{r})=\eta_*$, where we used the $Z_2$ symmetry $\eta(-x)=\eta(x)$ that follows from the boundary conditions $\eta(\pm\tilde{r})=\eta_*$.

\medskip
Also, the translational symmetry along the $x$-axis implies the following conservation law:
\begin{align}
\frac{a\big(\eta(x)\big)}{\sqrt{1-\eta'(x)^2}}=a\big(\eta_0\big)\,,
\label{eq:conserve}
\end{align}
where $\eta_0:=\eta(0)$ is the conformal time at the turning point of the geodesic. This implies that $a(\eta(x))\leq a(\eta_0)$. In particular, $\eta_0>\eta_*$ if the universe is in the expansion phase at $\eta=\eta_*$, which we assume in the following without loss of generality. Then, the geodesic is given by
\begin{align}
\label{tau-rho}
x(\eta)=\pm\int^{\eta_0}_{\eta}\frac{ d\tilde{\eta}}{\sqrt{1-\frac{a(\tilde{\eta})^2}{a(\eta_0)^2}}}\,,
\quad
\tilde{r}=|x(\eta_*)|=\int^{\eta_0}_{\eta_*}\frac{d\tilde{\eta}}{\sqrt{1-\frac{a(\tilde{\eta})^2}{a(\eta_0)^2}}}\,,
\end{align}
where the $\pm$ sign corresponds to the two halves of the geodesics. See also Figs.~\ref{fig:half-flat-d}--\ref{fig:half-flat-a2} for concrete profiles of geodesics.
Correspondingly, the geodesic distance reads
\begin{align}
\label{geoD}
D(\eta_*,\tilde{r})=2\int^{\eta_0}_{\eta_*}
d\eta\frac{a(\eta)^2}{\sqrt{a(\eta_0)^2-a(\eta)^2}}\,.
\end{align}
The holographic entanglement entropy is obtained by dividing the geodesic distance by $4G$. Our task is now to perform the integrals~\eqref{tau-rho}--\eqref{geoD}.

\paragraph{Analytic results for constant $w$.}

To be concrete, let us suppose that the universe is filled with an ideal fluid with a constant equation of state $w=p/\varepsilon$, where $\varepsilon>0$ and $p$ are the energy density and the  pressure of the fluid, respectively. Then, the scale factor $a(\eta)$ satisfies a simple power law (see also Appendix~\ref{scale} for basics of the FLRW universe):
\begin{align}
\label{scale-power}
a(\eta)=(w\eta)^{\frac{1}{w}}\,.
\end{align}
Here, we chose an expanding universe without loss of generality, so that the conformal time $\eta$ is in the range $0< \eta<\infty$ for the decelerating universe $w>0$ and $-\infty< \eta<0$ for the accelerating universe $w<0$, respectively. For the power-law scale factor~\eqref{scale-power}, the integrals~\eqref{tau-rho}--\eqref{geoD} are performed analytically:
\begin{align}
x(\eta)&=\pm\left[\eta_0\cdot\frac{\sqrt{\pi}\,\Gamma\left[\frac{w+2}{2}\right]}{\Gamma\left[\frac{w+1}{2}\right]}-\eta \cdot{}_2F_1\left[\frac{1}{2},\frac{w}{2};\frac{w+2}{2};\left(\frac{\eta}{\eta_0}\right)^{\frac{2}{w}}\right]\right]\,,
    \nn
    \\*
    \tilde{r}&=\eta_0 \cdot\frac{\sqrt{\pi}\,\Gamma\left[\frac{w+2}{2}\right]}{\Gamma\left[\frac{w+1}{2}\right]}-\eta _*\cdot{}_2F_1\left[\frac{1}{2},\frac{w}{2};\frac{w+2}{2};\left(\frac{\eta_*}{\eta_0}\right)^{\frac{2}{w}}\right]\,,
    \label{flatre}
    \\*
    D(\eta_*,\tilde{r})&=(w\eta_0)^{1+\frac{1}{w}}\frac{\sqrt{\pi}\,\Gamma\left[\frac{w+2}{2}\right]}{\Gamma\left[\frac{w+3}{2}\right]}
    -\frac{2}{w+2}
    \frac{(w\eta_*)^{1+\frac{2}{w}}}{(w\eta_0)^{\frac{1}{w}}}{}_2F_1\left[\frac{1}{2},\frac{w+2}{2};\frac{w+4}{2};\left(\frac{\eta_*}{\eta_0}\right)^{\frac{2}{w}}\right]\,,\nn
\end{align}
where each of the $\pm$ sign describes a half of the geodesics as we mentioned. Note that the above expression has an apparent singularity at $w=-2$ in each term, but they cancel each other to be finite. For instance, an explicit form of $\tilde{r}$ and $D(\eta_*,\tilde{r})$ for $w=-2$ is given by
\begin{align}
    \label{flatw2}
\tilde{r}=-\eta_*\sqrt{1-\frac{\eta_0}{\eta_*}}-\frac{\eta_0}{2}\ln\frac{1+\sqrt{1-\frac{\eta_0}{\eta_*}}}{1-\sqrt{1-\frac{\eta_0}{\eta_*}}}\,,
\quad
D(\eta_*,\tilde{r})&=\sqrt{-2\eta_0}\ln\frac{1+\sqrt{1-\frac{\eta_0}{\eta_*}}}{1-\sqrt{1-\frac{\eta_0}{\eta_*}}}
\,.
\end{align}
Also note that thanks to the power-law nature~\eqref{scale-power} of the scale factor, $\eta_*$-dependence of the geodesic length exhibits a simple power-law,
\begin{align}
\label{d_power}
D(\eta_*,\tilde{r})=|\eta_*|^{1+\frac{1}{w}}D\left(\pm 1,\frac{\tilde{r}}{|\eta_*|}\right)\,,
\end{align}
where the plus and minus signs are for the decelerating universe ($w>0$) and the accelerating universe ($w<0$), respectively.

\subsection{Half holography}

First, we apply the analytic results~\eqref{flatre} for the flat universe with constant $w$ to the half holography. Analogously to the de Sitter case, we place the boundary at a time $\eta=\eta_*$ on the straight line $y=0$ and take a subsystem on the interval,
\begin{align}
x\in (-r_*,r_*)\,,\quad y=0\,.
\end{align}
Then, the holographic entanglement entropy reads
\begin{align}
S=\frac{D(\eta_*,r_*)}{4G}\,.
\end{align}
The $\eta_*$-dependence of the entropy is a simple power-law (see Eq.\eqref{d_power})), which can be absorbed into an overall rescaling of the scale factor. Thus, we take $\eta_*=\pm1$ without loss of generality. In the following, we present the results for $w>0$, $-1<w<0$, and $w<-1$ in order.

\begin{figure}[t]
  \centering
\includegraphics[keepaspectratio, width=0.9\linewidth]{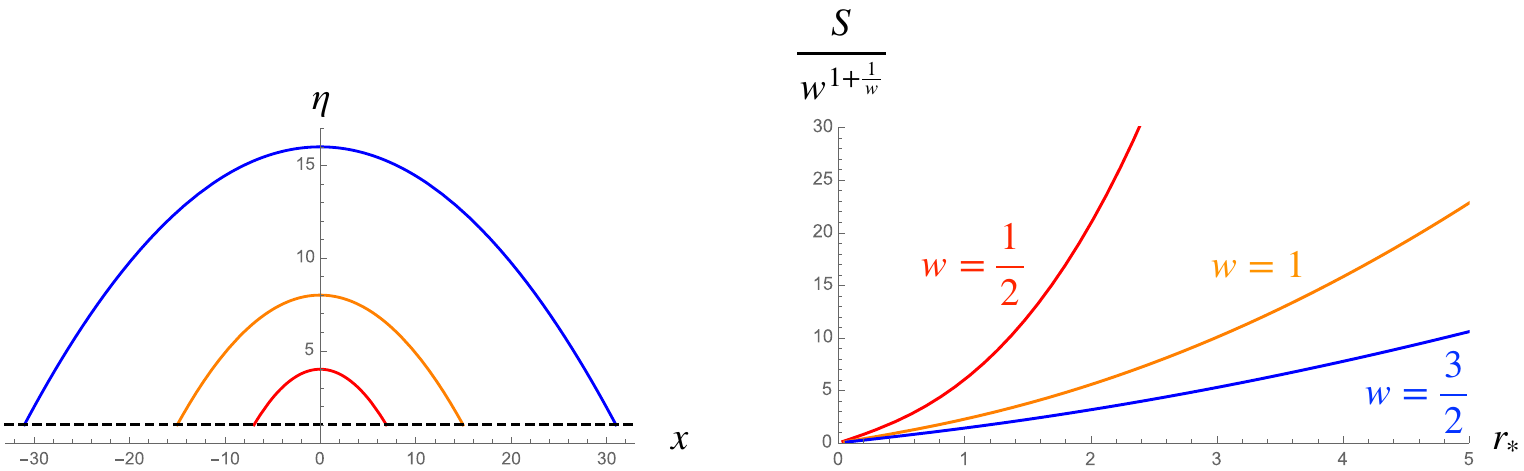}
\caption{Left: Profiles of geodesics for $w=\frac{1}{2}$. Red, orange, and blue curves are for $\eta_0=4,8,16$, respectively. The dashed line shows the $\eta_*=1$ slice, on which the subsystem is taken. Right: Holographic entanglement entropy for $w=\frac{1}{2},1,\frac{3}{2}$ in the unit $4G=1$. For visual clarity, $S/w^{1+\frac{1}{w}}$ is plotted in the figure.}
\label{fig:half-flat-d}
\end{figure}

\paragraph{Decelerating universes $(w>0)$.}

First, typical profiles of geodesics in the decelerating universe $(w>0)$ are shown in the left panel of Fig.~\ref{fig:half-flat-d}, indicating that for any $r_*$, there exists a single spacelike geodesic connecting the two boundary points. Indeed, as can be verified from Eq.~\eqref{flatre}, the subsystem size $r_*$ is a monotonic function of the turning time $\eta_0$ with an asymptotic behavior,
\begin{align}
r_*\simeq \eta_0 \cdot\frac{\sqrt{\pi}\,\Gamma\left[\frac{w+2}{2}\right]}{\Gamma\left[\frac{w+1}{2}\right]}
\quad
(\eta_0\gg\eta_*)\,.
\end{align}
Second, as shown in the right panel of Fig.~\ref{fig:half-flat-d}, the holographic entanglement entropy is a convex function of the subsystem size $r_*$. Hence, the half holography is incompatible with the SSA, similarly to the de Sitter case.

\begin{figure}[t]
  \centering
\includegraphics[keepaspectratio, width=0.95\linewidth]{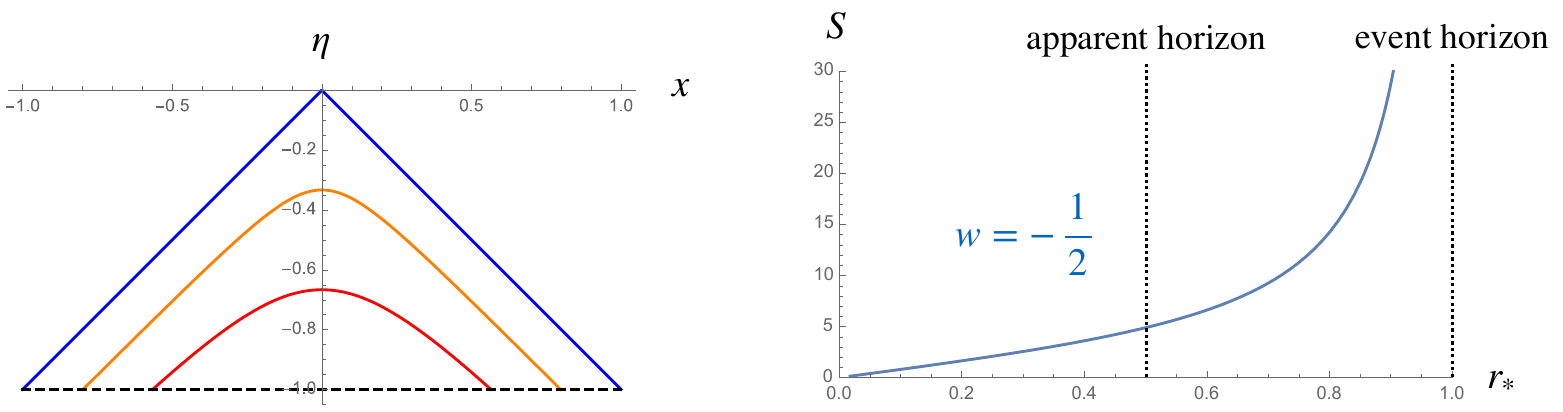}
\caption{Left: Profiles of geodesics for $w=-\frac{1}{2}$. Red, orange, and blue curves are for $\eta_0=-\frac{3}{4},-\frac{1}{2},0$, respectively. The dashed line shows the $\eta_*=-1$ slice, on which the subsystem is taken. Right: Holographic entanglement entropy for $w=-\frac{1}{2}$ in the unit $4G=1$. The entropy diverges when the subsystem touches the event horizon, whereas nothing special happens at the apparent horizon.}
\label{fig:half-flat-a1}
\end{figure}

\paragraph{Accelerating universes with $-1<w<0$.}

Next, we consider an accelerating universe that satisfies the null energy condition, i.e., $-1<w<0$, for which the apparent horizon $r=w\eta$ lies inside the event horizon $r=-\eta$ for an observer at the origin $r=0$. In particular, at $\eta=\eta_*=-1$, they are located at $r=|w|$ and $r=1$, respectively. As depicted in the left panel of Fig.~\ref{fig:half-flat-a1}, there exists a spacelike geodesic connecting the two boundary points only when the subsystem is inside the {\it event horizon} $r_*<1$. Accordingly, the holographic entanglement entropy is no more real when the subsystem stretches outside the event horizon. This is similar to the de Sitter case. Second, as shown in the right panel of Fig.~\ref{fig:half-flat-a1}, the holographic entanglement entropy is a convex function of the system size $r_*$. Hence, the half holography is again incompatible with the SSA. Note that the holographic entanglement entropy diverges when the subsystem touches the event horizon in contrast to the de Sitter case, as can be confirmed from the analytic formulae~\eqref{flatre} that the entropy asymptotically behaves as $S\propto (1-r_*)^{1+\frac{1}{w}}$ in the limit $r_*\to 1$.

\paragraph{Accelerating universes with $w<-1$.}

\begin{figure}[t]
  \centering
\includegraphics[keepaspectratio, width=1\linewidth]{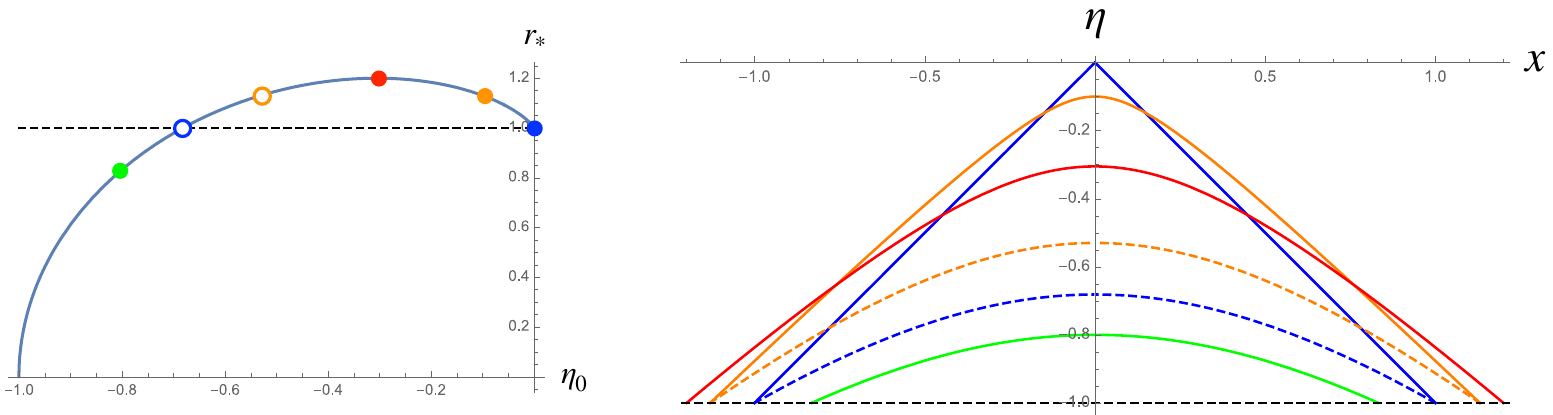}
\caption{Left: Subsystem size $r_*$ as a function of turning  time $\eta_0$ for $w=-2$. The dashed line shows the event horizon size $r_*=1$. The colored and white circles correspond to the solid and dashed curves of the same color in the right panel, respectively.
\\
Right: Profiles of geodesics for $w=-2$. The solid and doted curves in the same color have the same subsystem size $r_*$, and the solid one has a shorter geodesic length. Red, orange, blue and green are for $r_*\simeq 1.20,1.13,1.00,0.83$, respectively.}
\label{fig:half-flat-a2}
\end{figure}

Finally, we consider an accelerating universe that violates the null energy condition, i.e., $w<-1$, for which the apparent horizon $r=w\eta$ lies outside of the event horizon $r=-\eta$ for an observer at the origin $r=0$. The left panel of Fig.~\ref{fig:half-flat-a2} illustrates the subsystem size $r_*$ as a function of the turning time $\eta_0$, demonstrating that there exists a spacelike geodesic connecting the two boundary points, even when the subsystem stretches outside the event horizon $r=1$. More remarkably, there exist two distinct spacelike geodesics when $r_*$ is in the range $1\leq r_*< r_{\rm max}$, where $r_{\rm max}$ is the maximum subsystem size that accommodates a spacelike geodesic. These features are in sharp contrast to the accelerating universe satisfying the null energy condition $-1\leq w<0$. See also the right panel of Fig.~\ref{fig:half-flat-a2} for typical profiles of geodesics.

\begin{figure}[t]
  \centering
\includegraphics[keepaspectratio, width=0.55\linewidth]{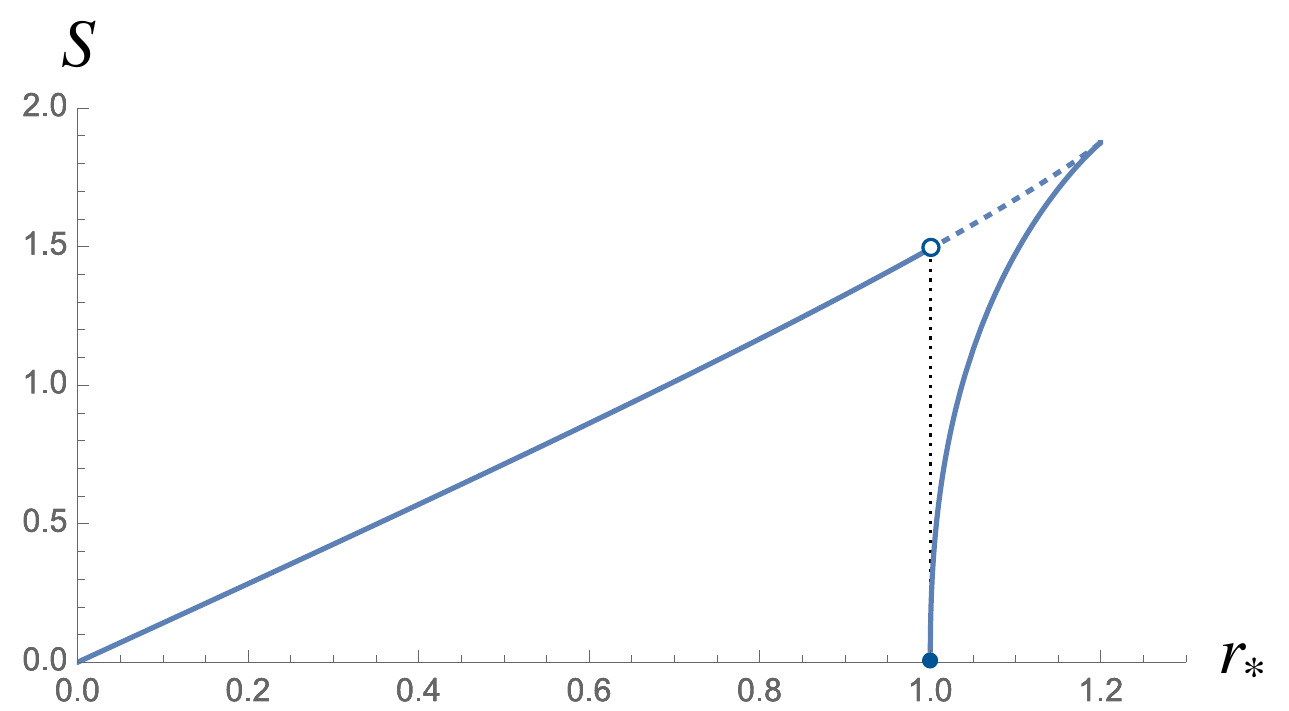}
\caption{
Holographic entanglement entropy for $w=-2$ as a function of subsystem size $r_*$.}
\label{fig:half-flat-a3}
\end{figure}

\medskip
Accordingly, the holographic entanglement entropy exhibits a peculiar phase transition at the event horizon scale $r_*=1$. See Fig.~\ref{fig:half-flat-a3}. When the subsystem lies inside the event horizon $r_*<1$, there exists a single geodesic for each $r_*$. In this regime, the holographic entanglement entropy is a convex function of $r_*$. At $r_*=1$, a new branch emerges, depicted by the solid curve in the regime $r_*\geq 1$ of the figure. This new branch has a shorter geodesic length than the original branch (the dotted curve) and therefore it is identified with the Ryu--Takayanagi geodesic. Note that the entropy is a concave function of $r_*$ in the regime $1\leq r_*\leq r_{\rm max}$ ($r_{\rm max}\simeq 1.20$ for $w=-2$ used in the figure). For $r_*>r_{\rm max}$, there is no spacelike geodesic connecting the two boundary points, so that the entropy is no more real. Evidently, the holographic entanglement entropy does not satisfy the SSA and hence the half holography is incompatible similarly to the previous examples. Since the null energy condition is violated, there may be a geodesic with real valued length even outside the event horizon.

\subsection{Horizon-type holography}
\label{FLRW:horizon}

Next, we study the horizon-type holography. Similarly to the de Sitter case, we place the boundary at a time $\eta=\eta_*$ on a circle defined by
\begin{align}
r=\lambda r_{\rm H}\,,
\quad
0\leq\phi<2\pi\,,
\end{align}
where $r_\text{H}:=\frac{a(\eta_*)}{a'(\eta_*)}$ is the radius of the {\it apparent horizon} and $\lambda$ parametrizes the boundary size compared to $r_\text{H}$. Note that $r_\text{H}=w\eta_*$ for the power-law scale factor~\eqref{scale-power}. Also, we take a subsystem on the arc,
\begin{align}
r=\lambda r_{\rm H}\,,
\quad
\phi\in(0,\phi_*)\,.
\end{align}
Then, $\tilde{r}$ in Eq.~\eqref{after_rotation} is
$\tilde{r}=\lambda r_{\rm H}\sin\frac{\phi_*}{2}$, and the holographic entanglement entropy reads
\begin{align}
S=\frac{D(\eta_*,\lambda r_{\rm H}\sin\frac{\phi_*}{2})}{4G}\,.
\end{align}
In the following, we take $\eta_*=\pm1$ without loss of generality and present the results for decelerating universes $w>0$ and accelerating universes $w<0$ in order.

\paragraph{Decelerating universes.}

\begin{figure}[t]
  \centering
\includegraphics[keepaspectratio, width=0.55\linewidth]{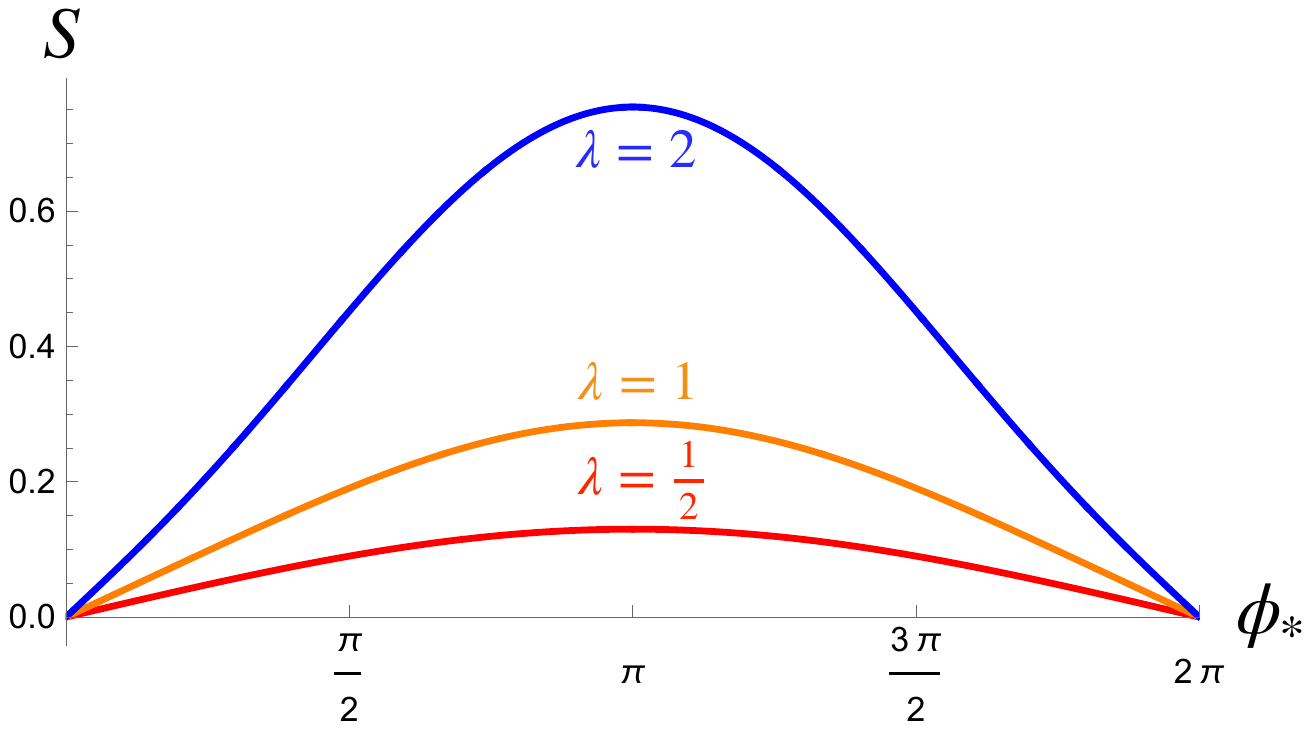}
\caption{
Holographic entanglement entropy for $w=\frac{1}{2}$ as a function of $\phi_*$. The boundary size is set as $\lambda=\frac{1}{2},1,2$, respectively.}
\label{fig:horizon-flat-de}
\end{figure}

Fig.~\ref{fig:horizon-flat-de} illustrates the holographic entanglement entropy for $w=\frac{1}{2}$ as a function of the subsystem size $\phi_*$, whose properties are summarized as follows: First, for any $w>0$, the entropy is real for any $\phi_*$. Second, it is concave when the subsystem is large enough. This is due to the $Z_2$ symmetry $\phi_*\to2\pi-\phi_*$ and the fact that the entropy is real for any $\tilde{r}$. However, if the boundary lies outside the apparent horizon $\lambda>1$, the entropy becomes convex in the small $\phi_*$ regime (see Sec.~\ref{subsub:short} for an analytic study). Therefore, the horizon-type holography is compatible with the SSA only when the holographic boundary is on or inside the apparent horizon.

\begin{figure}[t]
  \centering
\includegraphics[keepaspectratio, width=0.55\linewidth]{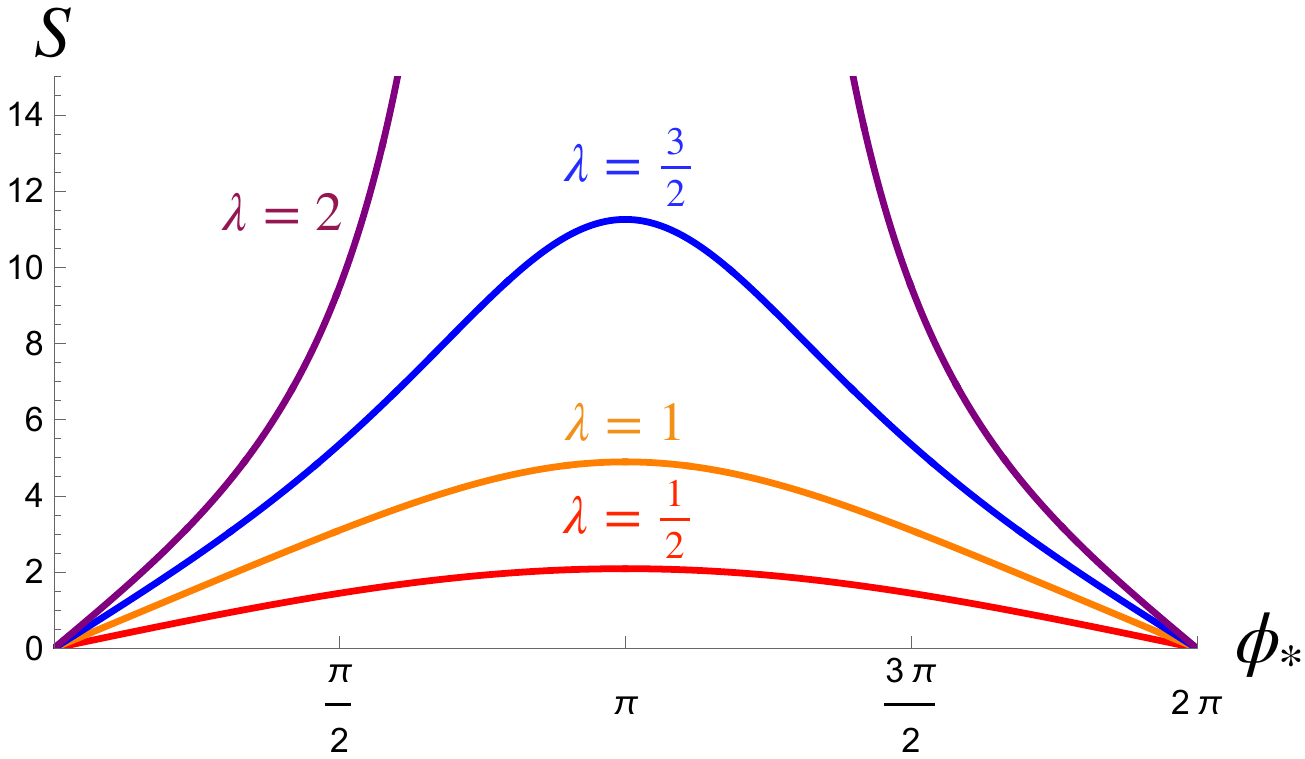}
\caption{
Holographic entanglement entropy for $w=-\frac{1}{2}$ as a function of $\phi_*$. The boundary size is set as $\lambda=\frac{1}{2},1,\frac{3}{2},2$, respectively.}
\label{fig:horizon-flat-a1}
\end{figure}

\begin{figure}[t]
  \centering
\includegraphics[keepaspectratio, width=0.55\linewidth]{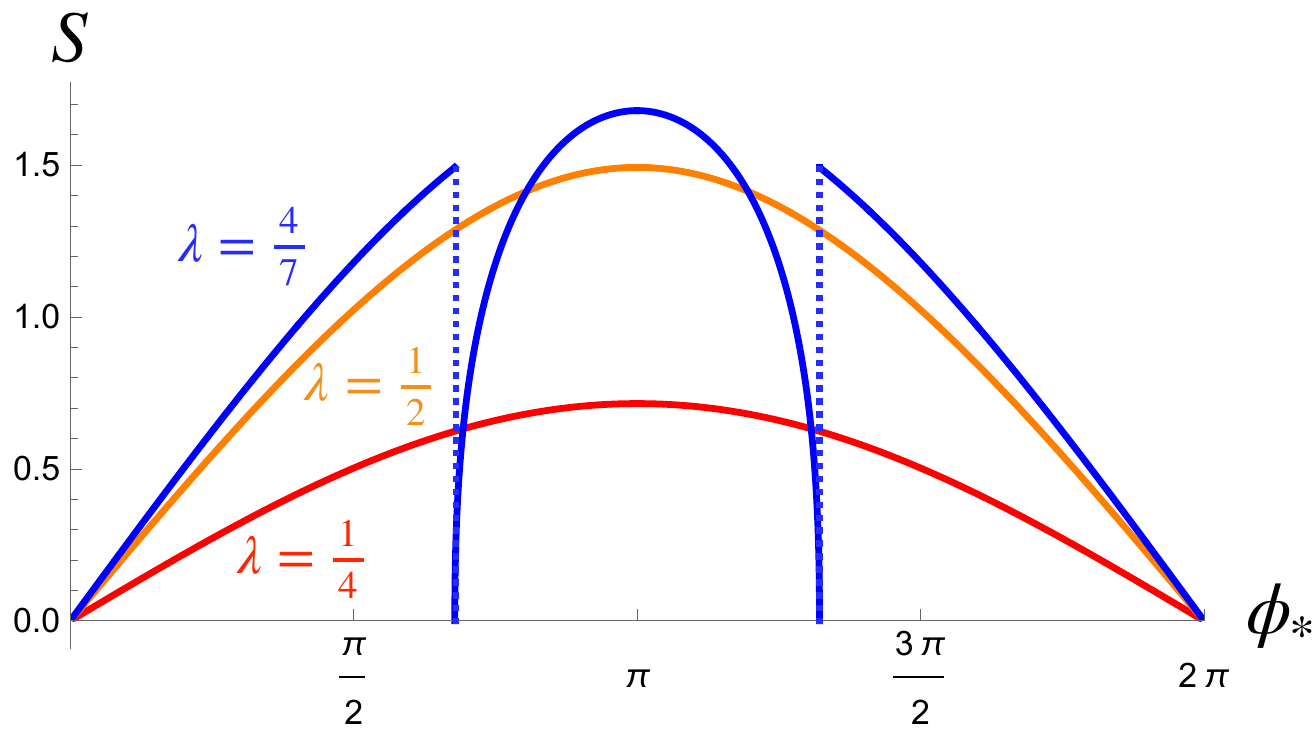}
\caption{
Holographic entanglement entropy for $w=-2$ as a function of $\phi_*$. The boundary size is set as $\lambda=\frac{1}{4},\frac{1}{2},\frac{4}{7}$, respectively.}
\label{fig:horizon-flat-a2}
\end{figure}

\paragraph{Accelerating universes.}

For the accelerating universe, the event horizon also plays a role. For $-1\leq w<0$, the holographic entanglement entropy has a real value only when the subsystem fits inside the event horizon. Therefore, the condition for the entropy to be real for all $\phi_*$ is $\lambda \leq |w|^{-1}$. The entropy for $w=-\frac{1}{2}$ as a function of $\phi_*$ is plotted in Fig.~\ref{fig:horizon-flat-a1}, where the entropy is concave and thus the horizon-type holography is compatible with the SSA only when the holographic boundary is on or inside the apparent horizon (see also Sec.~\ref{subsub:short}). On the other hand, for $w<-1$, the entanglement entropy exhibits a peculiar phase transition when the holographic boundary is outside the event horizon $\lambda>|w|^{-1}$. See Fig.~\ref{fig:horizon-flat-a2}. This may also be related with a violation of the null energy condition. Similarly to the half holography case, this makes the horizon-type holography with $\lambda>|w|^{-1}$ incompatible with the SSA. In contrast, if the holographic boundary is on or inside the event horizon $\lambda\leq|w|^{-1}$, the entanglement entropy is concave and compatible with the SSA.

\subsection{Where can we place the holographic boundary?}
\label{subsec:where}

To summarize so far, in the flat FLRW universe with a power-law scale factor, we found that the half holography is incompatible with the SSA, whereas the compatibility in the horizon-type holography depends on both the boundary size $\lambda$ and the equation of state parameter $w$. The necessary conditions for the SSA to be satisfied in the horizon-type holography can be characterized in terms of the subsystem size as follows:
\begin{enumerate}

\item Small subsystem

The SSA is satisfied in the small subsystem regime $\phi_*\ll1$, {\it only when the holographic boundary is on or inside the apparent horizon $\lambda\leq1$}. This statement about small subsystems is independent of the equation of state $w$ and applies to both accelerating and decelerating universes, since local properties are insensitive to details of the cosmic history. Indeed, we show below analytically that this statement holds in general FLRW spacetimes, beyond the flat universe with a constant $w$. Also, it is noteworthy that the apparent horizon is defined locally, which explains why the short-distance behavior of the holographic entanglement entropy depends on the apparent horizon.

\item Large subsystem

On the other hand, the long-distance behavior is sensitive to the event horizon, which is defined based on global causal structure of the spacetime. Its impact on the horizon-type holography is significant, especially when the event horizon of the accelerating universe is inside the apparent horizon, i.e., $w<-1$. If the holographic boundary is outside the event horizon, the holographic entanglement entropy experiences a peculiar phase transition, which leads to the violation of the SSA. Therefore, we arrive at a necessary condition for the SSA: {\it the holographic boundary in the accelerating universe has to be on or inside the event horizon}.

It is also interesting to rephrase our finding as follows. If the holographic boundary is placed on the apparent horizon, the SSA is satisfied {\it only when the universe satisfies the null energy condition $w\geq-1$.} This is reminiscent of earlier works~\cite{Allais:2011ys,Callan:2012ip,Caceres:2013dma} which uncover a similar condition between the null energy condition and consistency of the holographic entanglement entropy in other spacetimes.
\end{enumerate}
This observation leads to the following general expectation on the FLRW holography: If we require that the dual theory is an ordinary quantum field theory that respects unitarity and locality, {\it the holographic boundary has to be placed inside the apparent horizon and the event horizon of a geodesic observer}. In particular, if we focus on a symmetric embedding of the holographic boundary into the FLRW universe, a natural holographic scenario is the horizon-type holography with $\lambda\leq1$.\footnote{
Note that for $w>0$, the holographic entanglement entropy is concave even for  $\lambda>1$, if the subsystem is sufficiently large. A somewhat artificial but possible resolution to ensure the compatibility with the SSA may be to introduce a UV cutoff in the dual theory and restrict the range of the subsystem size $\phi_*$.} To support this claim, in Sec.~\ref{subsub:short}, we study the short-distance behavior of the holographic entanglement entropy in general FLRW universes.

\medskip
Before proceeding, we briefly comment on the original dS/CFT correspondence and possible extensions to the FLRW universe. As we remarked at the end of Sec.~\ref{subsec:setup}, our analysis can be applied directly to holography with a spacelike boundary at a constant time $\eta$. In particular, if one identify the $y$ coordinate with a Euclidean time coordinate of the would-be dual theory, the entanglement entropy coincides with that in the half holography studied above. Since the SSA is always violated in this case, the dual theory will not be a standard QFT. Indeed, it is expected to be non-unitary since the bulk and boundary do not share the notion of time. While it is interesting to explore such a holography with a non-standard dual theory, the approach in this paper is different: we are exploring a criteria for holographic and cosmological scenarios by postulating that the dual quantum field theory is well behaved.

\subsubsection{Short-distance behavior in general FLRW universe.}
\label{subsub:short}

To conclude this section, we analytically investigate the short-distance behavior of the holographic entanglement entropy in general FLRW universes. Note that, in contrast, the long-distance behavior depends on the entire cosmic history and so it is hard to make a general study. 

\medskip
Consider a general flat FLRW universe and recall the general formulae~\eqref{eq:geodesic}--\eqref{eq:conserve} of the geodesic length. Combining them, the holographic entanglement entropy reads
\begin{align}
S=\frac{1}{4G}\int_{-\tilde{r}}^{\tilde{r}}\frac{a^2(\eta(x))}{a(\eta(0))}dx\,,
\end{align}
where $\tilde{r}=r_*$ for the half holography and $\tilde{r}_*=\lambda r_{\rm H}\sin\frac{\phi_*}{2}$ for the horizon-type holography. The small $\tilde{r}$ behavior reads
\begin{align}
\label{S_small_r1}
S=\frac{1}{4G}
\left[
2a(\eta_*)\tilde{r}-\frac{1}{3}a'(\eta_*)\eta''(\tilde{r})\,\tilde{r}^3+\mathcal{O}(\tilde{r}^5)
\right]\,,
\end{align}
where we used the boundary condition $\eta'(0)=0$ of the problem. Meanwhile, the equation of motion following from Eq.~\eqref{eq:geodesic} implies
\begin{align}
\label{S_small_r2}
\eta''(\tilde{r})=-\frac{a'(\eta_*)}{a(\eta_*)}=-\frac{1}{r_{\rm H}}\,.
\end{align}
Substituting this into Eq.~\eqref{S_small_r1}, we obtain
\begin{align}
\label{S_small_rtilde}
S=\frac{a_*}{4G}
\left[
2\tilde{r}+\frac{1}{3}\frac{\tilde{r}^3}{r_{\rm H}^2}+\mathcal{O}(\tilde{r}^5)
\right]\,.
\end{align}
In the half holography, we have $\tilde{r}=r_*$ with $r_*$ being a half of the subsystem size, hence the entropy is convex for any flat FLRW universe in the short-distance regime. In Appendix~\ref{app:open/closed}, we find similar results for closed and open universes as well. Therefore, we conclude that the half holography is incompatible with the SSA for any FLRW universe.

\medskip
In the horizon-type holography, we have $\tilde{r}=\lambda r_{\rm H}\sin\frac{\phi_*}{2}$, where $\lambda$ parametrizes the size of the holographic boundary relative to the apparent horizon size, and $\phi_*$ is the subsystem size. Substituting this into Eq.~\eqref{S_small_rtilde} gives the small $\phi_*$ expansion:
\begin{align}
S=\frac{\lambda a_*r_{\rm H}}{4G}
\left[
\phi_*-\frac{1}{24}\left(1-\lambda^2\right)\phi_*^3+\mathcal{O}(\phi_*^5)
\right]\,.
\end{align}
As shown in Appendix~\ref{app:open/closed}, this result holds for both closed and open universes as well. We thus conclude that the horizon-type holography is compatible with the SSA only when the holographic boundary lies on or inside the apparent horizon $\lambda \leq1$. This supports our claim that the holographic boundary has to located inside the apparent horizon in order to have a standard dual QFT.

\section{Comments on matter entropy}\label{gee}

In our analysis so far, we have ignored the contributions of matter entropy to the holographic entanglement entropy. In this section, we argue that this approximation is naturally justified as long as the fluid description of the matter content in the universe is valid.

\medskip
For simplicity, let us assume that the subsystem is smaller than its complement (e.g., $0<\phi_*<\pi$ in the horizon-type holography). Then, the geometric contribution $S_{\rm geo}$, i.e., the length of the geodesic is estimated as
\begin{align}
S_{\rm geo}\sim \frac{L}{G}\,,
\end{align}
in terms of the physical size $L$ of the subsystem (e.g., $L=\lambda r_{\rm H}\phi_*$ in the horizon-type holography). On the other hand, the matter contribution corresponds to the matter entropy enclosed by the subsystem and the RT surface (which is simply a geodesic), and is estimated as
\begin{align}
S_{\rm mat}\sim sL^2,
\end{align}
where $s$ denotes the entropy density of the fluid. Therefore, the matter contribution is negligible as long as the subsystem size $L$ is sufficiently small:
\begin{align}
L\ll L_*\quad
{\rm with}
\quad
L_*\sim\frac{1}{Gs}\,.
\end{align}
As we showed in the previous section, the short-distance behavior of the holographic entanglement entropy is sufficient to conclude that the half holography and the horizon-type holography with $\lambda>1$ are incompatible with the SSA. Hence, regardless of the specific value of the critical length $L_*$, this conclusion remains unaffected by the matter entropy, since we can always choose $L$ to be much smaller than a given $L_*$.

\medskip
Next, let us examine the critical length $L_*$, in order to clarify applicability of our analysis to a long-distance subsystem. For concreteness, let us consider a flat universe filled with an ideal fluid. The entropy density follows from the standard thermodynamics as
\begin{align}
s=\frac{\varepsilon +p}{T}\,,
\end{align}
where we assumed a single-component fluid with temperature $T$ for simplicity. If the equation of state $w=p/\varepsilon$ satisfies $w+1=\mathcal{O}(1)$, we have $s\sim\varepsilon/T$, and thus the critical length reads
\begin{align}
L_*\sim \frac{T}{G\varepsilon}\,.
\end{align}
On the other hand, the Einstein equation relates the energy density $\varepsilon$ to the spacetime curvature as $G\varepsilon\sim H^2$ in terms of the Hubble parameter $H$. In addition, the physical size of the apparent horizon $R_{\rm H}=ar_{\rm H}$ is $R_{\rm H}=H^{-1}$. To sum up, the condition $L\ll L_*$ is rephrased as
\begin{align}
\label{LTH}
\frac{L}{R_{\rm H}}\ll \frac{T}{H}\,.
\end{align}
We emphasize that the matter entropy is of order $\mathcal{O}(G^{-1})$, which is at the same order as the leading geometric contribution. This is why the Newton constant $G$ does not appear in the condition~\eqref{LTH}. For instance, in the horizon-type holography with $\lambda=\mathcal{O}(1)$, we have $L_*/R_{\rm H}\lesssim 1$. Hence, the matter contribution is negligible as long as the fluid temperature is higher than the Hubble scale:
\begin{align}
T \gg H\,,
\end{align}
which is naturally satisfied as long as the fluid temperature is well-defined in the local patch of the universe. We thus conclude that the matter entropy is negligible and our analysis is applicable at least in a wide range of cosmological scenarios.

\medskip
As a caveat, it is worth mentioning that in realistic cosmological settings, the matter sector typically consists of multiple fluid components with distinct temperatures. Importantly, the dominant component of the entropy is not necessarily the same as the dominant component of the energy density, which invalidates the simplified single-field estimate discussed above. For example, in our universe, radiations such as the cosmic microwave and neutrino backgrounds is negligible in the current energy density, but they are important sources of entropy. In addition, dark matter halos, as well as Bekenstein--Hawking entropy associated with black holes and the cosmological event horizon, are investigated as the possible leading source of entropy in the universe~\cite{Egan:2009yy,Profumo:2024hnn}. It is interesting to explore how the previous conclusions are modified in scenario where the dominant source of matter entropy is  negligible as a source of the energy density. We leave a detailed analysis of such multi-component systems for future work.

\section{Conclusion}\label{con}
In this paper, we investigated the Ryu--Takayanagi surface, or the geodesic, in the three-dimensional FLRW universe. We considered two classes of holographic scenario called the half holography and the horizon-type holography in the bulk spacetime and defined a subsystem on a constant time slice. To examine possible location of boundaries, we utilized the strong subadditivity of the entanglement entropy.
Our analysis revealed that the strong subadditivity is generically violated in the half holography, whereas it can be satisfied in the horizon-type holography, provided that the holographic boundary lies on or inside the apparent horizon. Furthermore, in the universe filled with an ideal fluid satisfying a constant equation of state $w<-1$, this condition is strengthened: the holographic boundary has to be located inside the event horizon instead. In other words, assuming the null energy condition $w>-1$, a natural holographic scenario is the horizon-type holography with the holographic boundary placed on or inside the apparent horizon.

\section*{Acknowledgment}
We thank T.Takayanagi and S.M.Ruan for valuable discussions.
T.N. is supported in part by JSPS KAKENHI Grant No. JP22H01220 and MEXT KAKENHI Grant No. JP21H05184 and No. JP23H04007.
FS acknowledges financial aid from the Institute for Basic Science under the project code IBS-R018-D3, and JSPS Grant-in-Aid for Scientific Research No. 23KJ0938. 
YS is supported by Grant-in-Aid for JSPS Fellows No.\ 23KJ1337. We also thank the Chat GPT for revising the manuscript.
\appendix

\clearpage

\section{Basics of FLRW universe}\label{scale}

Here, we summarize the basics of the FLRW universe, including the closed and open universes.

\paragraph{Coordinates.}

The FLRW universe in $({d}{+}{1})$ dimensions is described by the metric,
\begin{align}
    ds^2&=a^2(\eta)
\left[
-d\eta^2+\frac{dr^2}{1-kr^2}+r^2d\Omega_{d-1}^2
\right]\,,
\end{align}
where $k=0,1,-1$ for the flat, closed, and open universes, respectively, and $r$ denotes the comoving radius of $S^{d-1}$. In particular, $d\eta=d\Omega_{d-1}=0$ is a great circle of the $d$-dimensional sphere and hyperboloid describing the closed and open universes. To manifest the rotational symmetry of the great circle, it is convenient to introduce a coordinate $\rho$ defined by
\begin{align}
\text{flat}:r=\rho
\,,
\quad
\text{closed}:r=\sin\rho
\,,
\quad
\text{open}:
r=\sinh\rho\,,
\label{eq:flrw_r_rho}
\end{align}
in terms of which the metric becomes
\begin{align}
    ds^2&=a^2(\eta)
\left[
-d\eta^2+d\rho^2+r^2(\rho)d\Omega_{d-1}^2
\right]\,.
\end{align}
In the following, we employ both coordinates systems interchangeably, depending on the context.
Note that $r\in[0,1]$ and $\rho\in[0,\pi]$ for the closed universe, whereas $r,\rho\in[0,\infty)$ for the flat and open universes. 

\paragraph{Apparent horizon.}

On the apparent horizon, the ingoing/outgoing light rays do not experience the expansion/contraction of the universe, respectively. This condition can be formulated by requiring that the time derivative of the physical radius vanishes for a null trajectory,
\begin{align}
    0=\frac{d}{d\eta}(ar)=r\frac{da}{d\eta}+a\frac{dr}{d\eta}=ra'\mp a\sqrt{1-kr^2}\,,
    \label{eq:apparent_horizon}
\end{align}
where the minus (plus) sign corresponds to the ingoing (outgoing) light ray in the expanding (contracting) universe. It leads to the radius $r_{\rm H}(\eta)$ of the apparent horizon at the time $\eta$,
\begin{align}
\label{r_H_general}
r_{\rm H}(\eta)=\frac{1}{\sqrt{\left (\frac{a'(\eta)}{a(\eta)}\right)^2+k}}\,.
\end{align}
The corresponding $\rho$ coordinate is written as $r_{\rm H}=\rho_{\rm H}$ (flat), $r_{\rm H}=\sin\rho_{\rm H}$ (closed), and $r_{\rm H}=\sinh\rho_{\rm H}$ (open), or
\begin{align}
\label{rho_H_general}
\text{flat}:\rho_{\rm H}=\frac{a}{|a'|}\,,
\quad
\text{closed}:\sin\rho_{\rm H}=\frac{a}{|a'|}\,,
\quad
\text{open}:\sinh\rho_{\rm H}=\frac{a}{|a'|}\,.
\end{align}

\paragraph{Benchmark examples.}

We consider the universe filled with an ideal fluid of the constant equation of state $w=p/\varepsilon$. From the Einstein equation, the scale factor reads
\begin{align}
\label{eq:a}
\text{flat}:
a(\eta)=\left(\frac{\eta}{\gamma}\right)^\gamma
\,,
\quad
\text{closed}:
a(\eta)=\left(\sin\frac{\eta}{|\gamma|}\right)^\gamma
\,,
\quad
\text{open}:
a(\eta)=\left(\sinh\frac{\eta}{\gamma}\right)^\gamma
\,,
\end{align}
where the parameter $\gamma=\frac{2}{d(1+w)-2}$ is the same as the one defined in Eq.~\eqref{power-law_genreal}, and for simplicity, we suppressed an overall dimensionful factor that characterizes the physical size of the universe, similarly to the flat universe case. Furthermore, without loss of generality, we focus on the expansion phase for the flat and open universes, leading to the time domain $0<\eta<\infty$ for the decelerating universe ($\gamma>0$) and $-\infty<\eta<0$ for the accelerating universe ($\gamma<0$). In the closed universe, the expansion and the contraction phase co-exist. The $\rho$ coordinate satisfies $\rho\in[0,\pi]$, and the domain of conformal time is $\eta\in[0,|\gamma|\pi]$. See the following tables for qualitative features of the closed universe:
\begin{enumerate}
\item {Decelerating closed universe $\gamma>0,k=1$}
\begin{center}
\begin{tabular}{|c|c|c|c|}\hline    $\eta=0$& $0<\eta<\gamma{\pi}/{2} $ &  $\gamma{\pi}/{2} <\eta<\pi \gamma $ &$\eta=\gamma\pi $  \\\hline Big Bang & expansion $\dot{a}>0$  & contraction $\dot{a}<0$& Big Crunch \\\hline  \end{tabular}
\end{center}
\item {Accelerating closed universe $\gamma<0,k=1$}
\begin{center}
\begin{tabular}{|c|c|}
\hline
$0<\eta<|\gamma|\pi/2 $
&
$|\gamma|{\pi}/{2}<\eta<|\gamma|\pi$
\\\hline
contraction $\dot{a}<0$
&
expansion $\dot{a}>0$
\\\hline
\end{tabular}
\end{center}
\end{enumerate}
From Eqs.~\eqref{r_H_general}--\eqref{rho_H_general}, the location of the apparent horizon in the $r$ coordinate is
\begin{align}
\text{flat}:
r_{\rm H}=\frac{\eta}{\gamma}\,,
\quad
\text{closed}:r_{\rm H}=\sin\frac{\eta}{|\gamma|}\,,
\quad
\text{open}:r_{\rm H}=\sinh\frac{\eta}{\gamma}\,,
\end{align}
or equivalently in the $\rho$ coordinate,
\begin{align}
\text{flat}:
\rho_{\rm H}=\frac{\eta}{\gamma}\,,
\quad
\text{closed}:\rho_{\rm H}=\left\{\begin{array}{cc}
    \dfrac{\eta}{|\gamma|} & (0< \eta\leq \frac{\pi}{2}) \\ 
    \pi-\dfrac{\eta}{|\gamma|} & \,\, (\frac{\pi}{2}\leq \eta< \pi)
    \end{array}\right.,
\quad
\text{open}:\rho_{\rm H}=\frac{\eta}{\gamma}\,.
\end{align}

\section{Closed and open universes}\label{app:open/closed}

In this appendix, we extend the analysis of the main text to the closed and open universes. 
The general FLRW universe in ${2}{+}{1}$-dimension is expressed as
\begin{align}
    ds^2=a^2(\eta)
\left[
-d\eta^2+\frac{dr^2}{1-kr^2}+r^2d\phi^2
\right]
=a^2(\eta)
\left[
-d\eta^2+d\rho^2+r^2(\rho)d\phi^2
\right]\,,
\end{align}
where $r(\rho)$ is given in Eq.~\eqref{eq:flrw_r_rho}, and $k=0,1,-1$ for the flat, closed, and open universes respectively. For a while, we focus on the formulation independent of the form of the scale factor.
See Appendix~\ref{scale} for more details of the spacetime.

\subsection{Geodesics in general FLRW universe}

Similarly to the flat universe case, we are interested in the geodesic distance between two points $A,B$ on the same constant-$\eta$ slice $\eta=\eta_*$:
\begin{align}
(\eta,\rho,\phi)=(\eta_*,\rho_A,\phi_A),\,\,(\eta_*,\rho_B,\phi_B)\,.
\end{align}
Using isometries of the FLRW spacetime, i.e., homogeneity and isotropy of the spatial slice $\eta=\eta_*$, we may map the two boundary points to
\begin{align}
\label{after_rotation_k}
(\eta,\rho,\phi)=(\eta_*,\tilde{\rho},0),\,\,(\eta_*,\tilde{\rho},\pi),
\end{align}
where $\tilde{\rho}$ is a half of the geodesic distance on the two-dimensional spatial slice $\eta=\eta_*$. Its explicit form reads
\begin{align}
\label{rho_*1}
\text{flat}&:
\,\,\,\,\,\,\,\,\,\,\, \tilde{\rho}\,=\frac{1}{2}\sqrt{\rho_A^2+\rho_B^2-2\rho_A\rho_B\cos(\phi_A-\phi_B)}\,,
\\
\label{rho_*2}
\text{closed}&:
\,\,\,\sin \tilde{\rho}\,=\sqrt{\frac{1-\cos\rho_A\cos\rho_B-\sin\rho_A\sin\rho_B\cos(\phi_A-\phi_B)}{2}}\,,
\\
\label{rho_*3}
\text{open}&:\,
\sinh \tilde{\rho}=
\sqrt{\frac{\cosh\rho_A\cosh\rho_B-1-\sinh\rho_A\sinh\rho_B\cos(\phi_A-\phi_B)}{2}}
\,.
\end{align}
The rotational symmetry implies that $d\phi=0$ along the geodesic Thus the problem reduces to the one in the two-dimensional spacetime $ds^2=a^2(\eta)[-d\eta^2+d\rho^2]$, which is nothing but the one we studied for the flat universe. Utilizing the results in the main text (see around Eqs.~\eqref{eq:conserve}--\eqref{geoD}), we find the following formula of the geodesic length $D(\eta_*,\tilde{\rho})$:
\begin{align}
\label{d-general-open/closed}
D(\eta_*,\tilde{\rho})=2\int^{\eta_0}_{\eta_*}
d\eta\frac{a(\eta)^2}{\sqrt{a(\eta_0)^2-a(\eta)^2}}
\quad
{\rm wtih}
\quad
\tilde{\rho}=\int^{\eta_0}_{\eta_*}\frac{d\tilde{\eta}}{\sqrt{1-\frac{a(\tilde{\eta})^2}{a(\eta_0)^2}}}\,,
\end{align}
where we assumed that the universe is in the expansion phase at $\eta=\eta_*$, so that the turning point $\eta_0$ of the geodesic satisfies $\eta_0>\eta_*$.

\paragraph{Short-distance behavior.}

Similarly to the flat universe case, the short-distance behavior of the geodesic length~\eqref{d-general-open/closed} can be analyzed without specifying details of the cosmic history. By performing the small $\tilde{\rho}$ expansion and using the equation of motion as well as the boundary conditions (see discussion around Eqs.~\eqref{S_small_r1}--\eqref{S_small_r2} for details), we find
\begin{align}
\label{general_short1}
D(\eta_*,\tilde{\rho})=
a(\eta_*)
\left[
2\tilde{\rho}+\frac{1}{3}\left(\frac{a'(\eta_*)}{a(\eta_*)}\right)^2\tilde{\rho}^3+\mathcal{O}(\tilde{\rho}^5)
\right]
\,.
\end{align}
It is also convenient to express it in terms of the $r$ coordinate defined by
\begin{align}
\label{tilder_general}
\text{flat}:\tilde{r}=\tilde{\rho}
\,,
\quad
\text{closed}:\tilde{r}=\sin\tilde{\rho}
\,,
\quad
\text{open}:\tilde{r}=\sinh\tilde{\rho}
\,,
\end{align}
in which case, the expansion becomes
\begin{align}
\label{general_short2}
D(\eta_*,\tilde{\rho})=
a(\eta_*)
\left[
2\tilde{r}+\frac{1}{3}
\frac{\tilde{r}^3}{r_{\rm H}^2}+\mathcal{O}(\tilde{r}^5)
\right]
\,,
\end{align}
where $r_{\rm H}$ is the radius of the apparent horizon at the time $\eta=\eta_*$, given in Eq.~\eqref{r_H_general}. In the following subsection, we discuss the short-distance behavior of the holographic entanglement entropy based on Eq.~\eqref{general_short1} and Eq.~\eqref{general_short2}.

\subsection{Holographic scenarios}

Similarly to the flat universe case, we consider the following two types of holographic scenarios. This subsection summarizes general properties of the holographic entanglement entropy, leaving detailed analyses in specific cosmological scenarios to the subsequent part of the section.

\begin{figure}[t]
  \centering
\includegraphics[keepaspectratio, width=0.7\linewidth]{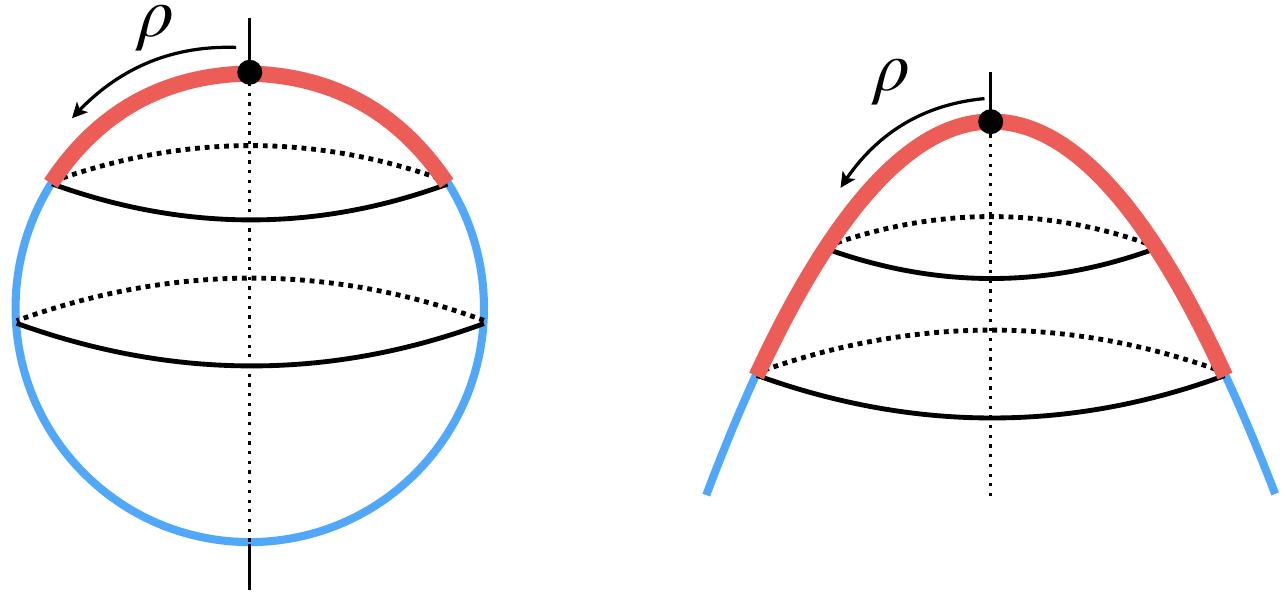}
\caption{
The two-dimensional sphere (left) and hyperboloid (right) describe the closed and open universes at $\eta=\eta_*$. The black point is for the North pole $\rho=0$ and the rotation angle around the straight line is $\phi$.
The subsystem and its complement in the half holography are depicted by the red and blue curves, respectively.}
\label{fig:half-open_closed}
\end{figure}

\paragraph{Half holography.}

In the half holography, we place the holographic boundary at $\eta=\eta_*$ on a curve defined by
$\rho\sin\phi=0$,
which bisects the universe. Note that the shift symmetry along the holographic boundary is guaranteed by the isometry. As a subsystem, we consider an interval defined by
\begin{align}
\rho\cos\phi\in[-\rho_*,\rho_*]
\,,
\quad
\rho\sin\phi=0\,.
\end{align}
See also Fig.~\ref{fig:half-open_closed}. Then, the holographic entanglement entropy reads
\begin{align}
S=\frac{D(\eta_*,\rho_*)}{4G}\,.
\end{align}
In particular, its short-distance behavior follows from Eq.~\eqref{general_short1} as
\begin{align}
S=\frac{a(\eta_*)}{4G}
\left[
2\tilde{\rho}+\frac{1}{3}\left(\frac{a'(\eta_*)}{a(\eta_*)}\right)^2\tilde{\rho}^3+\mathcal{O}(\tilde{\rho}^5)
\right]\,,
\end{align}
which shows that the entropy is always convex in the short-distance regime. Therefore, we conclude that the half holography is incompatible with the SSA in the closed and open universes as well.

\begin{figure}[t]
  \centering
\includegraphics[keepaspectratio, width=0.7\linewidth]{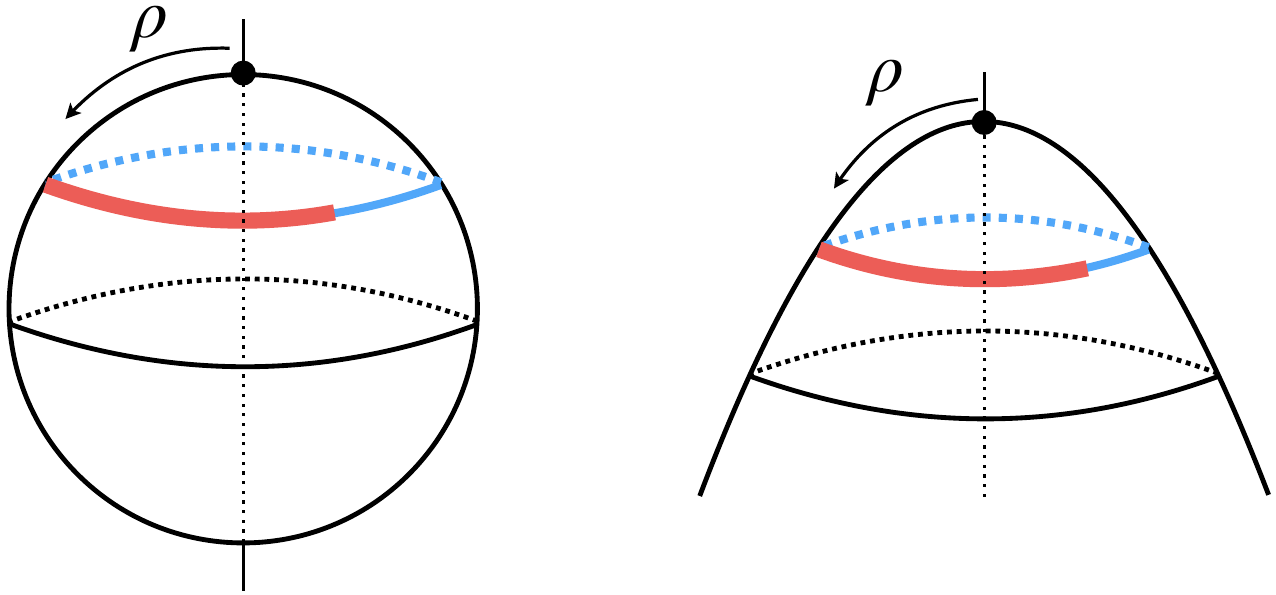}
\caption{The two-dimensional sphere (left) and hyperboloid (right) describe the closed and open universes at $\eta=\eta_*$.
The subsystem and its complement in the horizon-type holography are depicted by the red and blue curves, respectively.}
\label{fig:horizon-open_closed}
\end{figure}

\paragraph{Horizon-type holography.}

In the horizon-type holography, the apparent horizon plays an important role, whose radius $r_{\rm H}=r_{\rm H}(\eta_*)$ at the time $\eta=\eta_*$ is given in Eq.~\eqref{r_H_general}.
We place the holographic boundary on a circle defined in the $r$ coordinate by
\begin{align}
r=\lambda r_{\rm H}
\end{align}
or equivalently, in the $\rho$ coordinate,
\begin{align}
\text{flat}:\rho=\lambda r_{\rm H} \,,
\quad
\text{closed}: \sin\rho=\lambda r_{\rm H} \,,
\quad
\text{open}:
\sinh\rho=\lambda r_{\rm H}\,,
\end{align}
where $\lambda$ quantifies the size of the boundary relative to the apparent horizon. As a subsystem, we consider an arc defined by
\begin{align}
r=\lambda r_{\rm H}\,,
\quad\phi\in[0,\phi_*]\,.
\end{align}
and similarly in the $\rho$ coordinate. See also Fig.~\ref{fig:horizon-open_closed}.
The corresponding $\tilde{\rho}$ defined in Eqs.~\eqref{rho_*1}--\eqref{rho_*3} reads
\begin{align}
\text{flat}:
\tilde{\rho}=\lambda r_{\rm H}\sin\frac{\phi_*}{2}
\,,
\quad
\text{closed}:
\sin\tilde{\rho}=\lambda r_{\rm H}\sin\frac{\phi_*}{2}
\,,\quad
\text{open}:
\sinh\tilde{\rho}=\lambda r_{\rm H}\sin\frac{\phi_*}{2}\,.
\end{align}
In terms of $\tilde{r}$ defined in Eq.~\eqref{tilder_general}, the above relations are uniformly expressed as
\begin{align}
\tilde{r}=\lambda r_{\rm H}\sin\frac{\phi_*}{2}\,.
\end{align}
The holographic entanglement entropy then reads
\begin{align}
\text{flat}&:
S=\frac{D(\eta_*,\lambda r_{\rm H}\sin\frac{\phi_*}{2})}{4G}
\,,\\
\text{closed}&:
S=\frac{D(\eta_*,\arcsin[\lambda r_{\rm H}\sin\frac{\phi_*}{2}])}{4G}
\,,\\
\text{open}&:
S=\frac{D(\eta_*,\text{arsinh}[\lambda r_{\rm H}\sin\frac{\phi_*}{2}])}{4G}\,.
\end{align}
The short-distance behavior follows from Eq.~\eqref{general_short2} as
\begin{align}
S=\frac{\lambda a_*r_{\rm H}}{4G}
\left[
\phi_*-\frac{1}{24}\left(1-\lambda^2\right)\phi_*^3+\mathcal{O}(\phi_*^5)
\right]\,.
\end{align}
Hence, in the short-distance regime, the horizon-type holography is compatible with SSA, only when the holographic boundary is on or inside the apparent horizon.

\subsection{de Sitter spacetime}
\label{subsec:closed/open}

As a concrete example, we consider the de Sitter spacetime, in which the holographic entanglement entropy can be computed analytically by evaluating the integral~\eqref{d-general-open/closed} or by using the embedding coordinate expression~\eqref{geodesic_length_dS_emb}. Here, we present the latter analysis and show that qualitative features of the entropy do not depend on the curvature $k$ of the universe.

\subsubsection{Coordinate systems}

We first introduce the closed and open charts of de Sitter spacetime and derive an analytic formula for the geodesic length using the embedding language.

\paragraph{Closed universe.}

The de Sitter closed universe is described by the coordinates,
\begin{align}
    X^0=-\frac{1}{\tan \eta},\quad X^1=\frac{\cos\rho}{\sin\eta},\quad X^2=\frac{\sin\rho}{\sin\eta}\cos\phi,\quad    X^3=\frac{\sin\rho}{\sin\eta}\sin\phi
\end{align}
with
\be
0< \eta< \pi\,,\quad 0\leq\rho\leq\pi\,,\quad 0\leq\phi< 2\pi\,.
\ee
The corresponding metric is
\be
ds^2=\frac{-d\eta^2+d\rho^2+\sin^2\rho \,d\phi^2}{\sin^2 \eta}
\,.
\ee
The cosmological event horizon for the observer sitting at $\rho=0$ is located at
\begin{align}
\rho=\left\{\begin{array}{cc}
\eta
& (0<\eta\leq\frac{\pi}{2})
\\
\pi-\eta
& (\frac{\pi}{2}\leq\eta <\pi)\end{array}\right..
\end{align}
The geodesic distance $D$ of two points $(\eta_A,\rho_A,\phi_A)$ and $(\eta_B,\rho_B,\phi_B)$ reads
\be
\cos D=1-\frac{\cos(\eta_A-\eta_B)-\cos\rho_A\cos\rho_B-\sin\rho_A\sin\rho_B\cos(\phi_A-\phi_B)}{\sin\eta_A\sin\eta_B}\,.
\ee

\paragraph{Open universe.}

The de Sitter open universe is described by the coordinates,
\begin{align}
    X^0=-\frac{\cosh\rho}{\sinh\eta}\,,\quad X^1=\frac{1}{\tanh\eta}\,,\quad X^2=\frac{\sinh\rho}{\sinh\eta}\cos\phi\,,\quad  X^3=\frac{\sinh\rho}{\sinh\eta}\sin\phi
\end{align}
with
\be
-\infty<\eta<0\,,\quad 0\leq\rho<\infty\,,\quad 0\leq\phi<2\pi\,.
\ee
The corresponding metric is
\be
ds^2=\frac{-d\eta^2+d\rho^2+\sinh^2\rho \,d\phi^2}{\sinh^2\eta}\,.
\ee
The cosmological event horizon for the observer sitting at $\rho=0$ is located at $\rho=-\eta$. The geodesic distance of two points $(\eta_A,\rho_A,\phi_A)$ and $(\eta_B,\rho_B,\phi_B)$ reads
\be
\cos D=1-\frac{-\cosh(\eta_A-\eta_B)+\cosh\rho_A\cosh\rho_B-\sinh\rho_A\sinh\rho_B\cos(\phi_A-\phi_B)}{\sinh\eta\sinh\eta_B}\,.
\ee

\subsubsection{Holographic entanglement entropy}

Now, we compute the holographic entanglement entropy in the two holographic scenarios. 

\paragraph{Half holography.}

In the half holography, we consider the subsystem at the time $\eta=\eta_*$ as
\begin{align}
\rho\cos\phi\in[-\rho_*,\rho_*]\,,
\quad
\rho\sin\phi\,.
\end{align}
Then, the holographic entanglement entropy $S=\frac{D}{4G}$ with $D$ being the geodesic distance of the two points on the boundary reads
\begin{align}
\text{closed}:
\,\,
S=\frac{1}{2G}\arcsin \frac{\sin \rho_*}{\sin\eta_*}\,,
\quad
\text{open}:\,\,
S=\frac{1}{2G}\arcsin \frac{\sinh \rho_*}{\sinh\eta_*}\,,
\end{align}
which is real as long as the subsystem interval is inside the horizon. More explicitly, we may rewrite it in terms of the $r$ coordinate, $r_*=\sin\rho_*$ (closed) and $r_*=\sinh\rho_*$ (open), and the horizon radius, $r_{\rm H}=\sin\eta_*$ (closed) and $r_{\rm H}=\sinh\eta_*$ (open), as
\begin{align}
S=\frac{1}{2G}\arcsin \frac{r_*}{r_{\rm H}}\,.
\end{align}
This formula summarizes the results for all the flat, closed, and open universes. See also the flat universe result~\eqref{half-flat-dS}, where  $r_{\rm H}=1$. On the other hand, the second derivative of the entropy with respect to (half of) the subsystem size $\rho_*$ is
\begin{align}
\text{closed}&:
\,\,
\frac{\partial^2S}{\partial \rho_*^2}
=\frac{1}{2G}\frac{\cos^2\eta_*\sin\rho_*}{(\sin\eta_*^2-\sin\rho_*^2)^{\frac{3}{2}}}\,,
\\
\text{open}&:
\,\,
\frac{\partial^2S}{\partial \rho_*^2}
=\frac{1}{2G}\frac{\cosh^2\eta_*\sinh\rho_*}{(\sinh\eta_*^2-\sinh\rho_*^2)^{\frac{3}{2}}}\,.
\end{align}
Similarly to the flat universe case, the second derivative is always positive, hence the half holography is incompatible with the SSA.

\paragraph{Horizon-type holography.}

In the horizon-type holography, we take a subsystem at the time $\eta=\eta_*$ in the $r$ coordinate as
\begin{align}
r=\lambda r_{\rm H}
\,,
\quad\phi\in[0,\phi_*]
\end{align}
or equivalently in the $\rho$ coordinate as
\begin{align}
\text{closed}: \sin\rho=\lambda r_{\rm H} \,,\,\,\phi\in[0,\phi_*]\,,
\quad
\text{open}:
\sinh\rho=\lambda r_{\rm H}\,,\,\,\phi\in[0,\phi_*]\,.
\end{align}
Note that the location of the holographic boundary and the choice of the subsystem are essentially the same as the flat universe case. As a result, the holographic entanglement entropy and its second derivative take the same form as before:
\begin{align}
S=\frac{1}{2G}\arcsin\left[
\lambda\sin\frac{\phi_*}{2}
\right]\,,
\quad
\frac{\partial^2 S}{\partial\phi_*^2}
=-\frac{(1-\lambda^2)}{8G}\frac{ \lambda\sin\frac{\phi_*}{2}}{(1-\lambda^2\sin^2\frac{\phi_*}{2})^{\frac{3}{2}}}\,,
\end{align}
which are applicable for all the flat, closed, and open universes. We conclude also for the closed and open universes that the horizon-type holography is compatible with the SSA as long as the boundary is on or inside the horizon. 

\subsection{FLRW universe with constant equation of state}

\begin{figure}[t]
    \centering
    \includegraphics[width=0.45\linewidth]{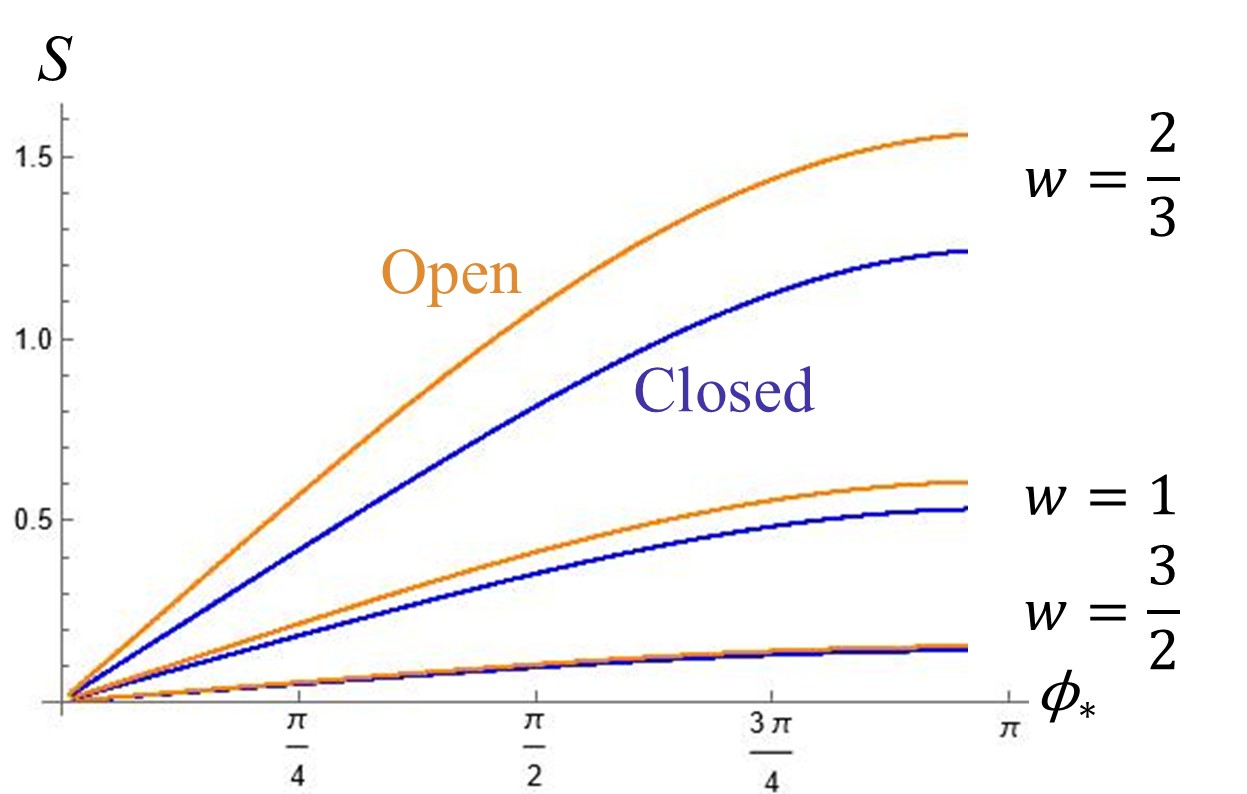}
    \includegraphics[width=0.45\linewidth]{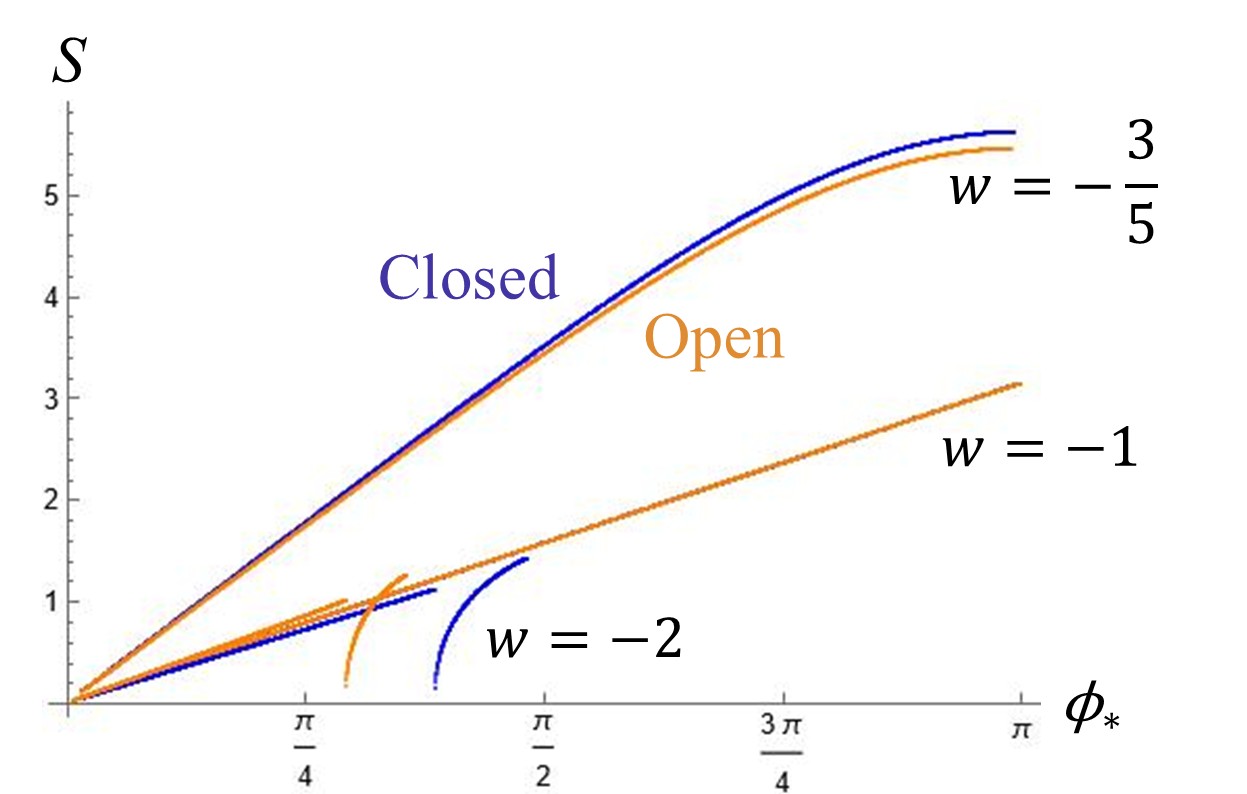}
    \caption{The holographic entanglement entropy in open universe (orange) and closed universe (blue). We fix the boundary at the apparent horizon $\lambda=1$ at the time slice $\eta_*=0.5$ for the decelerating universes (left) and $\eta_*=-0.5$ for the accelerating universes (right), respectively. Note that the plots exclude $\phi_*\simeq 0,\pi$ due to numerical error.}
    \label{fig:comp}
\end{figure}

Finally, we consider the closed and open universes $k=\pm 1$ filled with an ideal fluid of a constant equation of state parameter $w=p/\varepsilon$, where $\varepsilon>0$ and $p$ are the energy density and pressure of the fluid, respectively. In these cases, the scale factor $a(\eta)$ takes the form,
\begin{align}
\text{closed}:
a(\eta)=[\sin(w\eta)]^{\frac{1}{w}}
\,,
\quad
\text{open}:
a(\eta)=[\sinh(w\eta)]^{\frac{1}{w}}
\,.
\label{eq:a_open_close}
\end{align}
See also Eq.~\eqref{eq:a}.
Here, without loss of generality, we normalized the scale factor such that $a(\eta)=1$ when $\sin(w\eta)=1$ for the closed universe and $\sinh(w\eta)=1$ for the open universe. Regarding the domain of the coordinates, $0< w\eta<\pi$ for the closed universe, and we choose the expanding universe for open universe, leading to $0<\eta<\infty$ for the decelerating universe $w>0$ and $-\infty<\eta<0$ for the accelerating universe $w<0$. By substituting the scale factor into (\ref{d-general-open/closed}), the expressions for the RT surface are obtained. Although the integrals cannot be evaluated analytically, we perform the numerical analysis for the horizon-type boundary. The results are illustrated in Fig.~\ref{fig:comp}, which show that the qualitative features of the holographic entanglement entropy are insensitive to the curvature $k$. In particular, the horizon-type holography is compatible with the SSA as long as the boundary is on or inside the horizon, supporting the conclusion in the main text. 
The difference of the open and closed universe is their expansion rate~\eqref{eq:a_open_close}. In the decelerating universe, the open universe exhibits a larger physical volume at a given time and equation of state parameter, resulting in a larger RT surface. In contrast, in the accelerating universe with $w>-1$, the closed universe is physically larger, leading to a larger RT surface.
As a special case, the holographic entanglement entropy in de Sitter space $w=-1$ is the same for all $k$, in agreement with the conclusion in the previous subsection.
Finally, we remark that the flat universe case interpolates between those of the open and closed universe.

\bibliographystyle{JHEP_old}
\bibliography{NSSpaper.bib}

\end{document}